\documentstyle[preprint,aps]{revtex}
\begin{document}
\title{The two-leg $t-J$ ladder: A mean-field description}
\author{Y.L. Lee$^a$, Y.W. Lee$^a$, C.-Y. Mou$^a$, and Z.Y. Weng$^b$}
\address{$^a$NCTS and Physics Department, National Tsing Hua University, \\
Hsinchu, Taiwan}
\address{$^b$Texas Center for Superconductivity, University of Houston,\\ 
Houston, TX 77204-5932, U.S.A.}
\date{\today}
\maketitle

\begin{abstract}
   Two-leg $t-J$ ladders are investigated in the framework of a combination of the phase 
string formulation and bond-operator representation. We develope a mean-field theory 
in the strong rung interaction regime, i.e. $J_{\perp }\gg J,t$, which provides a unified 
description of the undoped insulating phase and the low doping phase --- the so-called 
$C1S0$ phase. Both of them are characterized by the resonating-valence-bond (RVB) 
order parameter, with gap opened up in all spin excitations. The ground state of the doped 
phase is intrinsically a superconductor with a d-wave symmetry, which is driven by the 
RVB correlations. The ground-state energy is in good agreement with numerical results. 
Phase separation is shown to occur beyond some critical value of $J/t$ for given doping 
concentration. We also show that the spin gap in the doped phase is determined by 
quasi-particle-like excitations. The local structure of hole pairs as well as the spectra of 
various spin and charge modes are analyzed in comparison with other approaches.
\end{abstract}


%%%%%%%%%%%%%%%%%%%%%%%%%%%%%%%%%%%%%%%%%%%

\section{Introduction}

    Since the discovery of high-$T_c$ superconductors, the study of $t-J$ type models has 
become an important topic of strongly correlated electron systems. The $t-J$ model 
provides us one of the simplest examples of the nontrivial interplay between charge and 
antiferromagnetism in a doped Mott insulator. It is a problem generally difficult to tackle 
by conventional many-body methods due to the nature of strong correlations.
 
Recently, ladder systems have received intensive studies both theoretically
\cite{Dago96,Dagotto92,Barnes,Reig,Gopalan,Noack,Sigrist,Poil,Troyer,White,Sierra97,Muller,Ammon}
and experimentally\cite{Azuma,Ueha,Maya,Imai}. From theoretical point of view, the  
ladder $t-J$ systems may be easier to investigate both numerically and analytically than 
the two-dimensional (2D) case related to the high-$T_c$ cuprates. However, the former 
may already catch some key physics of the latter and offer some important insights into 
the competition between charge and spin correlations beyond one-dimensional (1D) 
geometry.

Like in 1D and 2D, the physics of ladder $t-J$ systems has been more or less
well-understood at half-filling\cite{Barnes,Reig,Gopalan,Noack} where only the spin 
degrees of freedom are present. For example, in the two-leg ladder problem, the ground 
state may be visualized as a spin liquid state of the condensate of short-ranged RVB spin 
singlets\cite{Noack}. In the strong rung interaction limit, a bond-operator 
representation\cite{Sachdev} provides a very useful description\cite{Gopalan} of such a 
spin liquid state.

Also similar to 1D and 2D, the doped case in ladder $t-J$ systems poses a real challenge 
to analytic approaches: The central issue is how to correctly handle the competing charge 
and spin correlations once holes are introduced. In the two-leg ladder system, it has been 
established on numerical basis that on the small doping side ($\delta < 0.5$) there exists 
a so-called $C1S0$ phase\cite{Poil,Troyer} where spin excitations are all gapped 
(denoted by $S0$) while the density fluctuations of holes represent the only gapless mode 
there (denoted by $C1$). Such a phase is a superconducting phase with a d-wave-like 
symmetry. With the increase of the ratio $J/t$ (where $t$ and $J$ are two parameters of 
the $t-J$ model to be defined later), eventually a phase separation is found beyond some 
critical value of $J/t$.

Previously, a mean-field theory of lightly doped two-leg ladders proposed by 
Sigrist {\it et al.}\cite{Sigrist} showed that the short-ranged RVB state evolves into a 
superconductor with modified d-wave symmetry. It also gave a continuous evolution of 
the spin gap with doping. However, this theory fails to capture many features of the 
ground state and the excitation spectrum uncovered later by numerical 
work\cite{Poil,Troyer,White,Sierra97,Muller,Ammon}. The main reason is that the 
fermionic RVB mean-field theory employed in this approach is not a very good 
description even at half-filling. It is the purpose of this paper to develop an analytic 
framework that can give a unified and systematic description of the aforementioned 
physical properties in both the undoped and doped phases of the two-leg ladder $t-J$ 
model. We shall start from the strong rung interaction limit where it is natural to adopt 
the previously mentioned bond-operator representation\cite{Gopalan,Sachdev} for the 
undoped case. The corresponding mean-field theory\cite{Gopalan} gives a reasonably 
good description of the ground state and spin excitations at half-filling as memtioned 
before. In the doped case, additional bond operators are apparently needed and they can be 
classified as the rung hole pair ($d$) and quasiparticles ($a_\sigma $ and 
$\bar{a}_\sigma $ ) besides the spin singlet ($s$) and spin triplet ($t_\alpha$ ) operators. 
They compose of a complete basis convenient for describing the low doping case.

For any dimensionality, it is highly nontrivial to get access to the doped phase from the 
undoped insulator of the $t-J$ model. This is due to the fact that 
the Marshall signs\cite{Marsh} hidden in the half-filled spin background will be generally 
``disordered'' by the motion of holes, leading to the phase string effect\cite{Weng}. Such a 
phase string effect cannot be repaired through spin flip processes as the latter always 
respects the Marshall sign at low energy. It implies that the nonrepairable ``phase strings'' 
left on the hole paths will be present in the ground state of the doped case. These 
``phase strings'' play a role similar to the Fermi-surface phase-shifts originally proposed by 
Anderson. Indeed, in 1D case, the phase string effect leads to the Luttinger liquid behavior. 
The phase string formulation\cite{Weng} provides a systematical method which reproduces 
correct exponents of various correlation functions. Nontrivial phase string effect in 2D 
mean-field theory has been also investigated in Ref.\cite{Weng98}. Incorporating the phase 
string effect thus becomes a necessary step to construct a sensible mean-field theory in the 
study of the two-leg ladder systems. 

Starting from the half-filling where the mean-field theory is based on a RVB 
characterization with an order parameter $\langle s_j\rangle =\bar{s}$, we are able to 
generalize the theory to the doped regime after incorporating the phase string effect. We 
find that the ground state in the doped phase is naturally a superconductor with a 
d-wave-like symmetry, as the consequence of the RVB correlations in the insulating 
phase. The mean-field ground-state energy is in good agreement with the numerical one 
at the doping concentration $\delta \leq 0.5$. Moreover, an instability of phase separation 
occurs in our mean-field state as the ratio $J/t$ increases beyond some critical value, also 
consistent with numerical results. The present mean-field theory thus accommodates the 
most important physical properties of the doped two-leg $t-J$ model previously identified 
only numerically. We would like to point out that without explicitly dealing with the 
nonlocal phase string effect at the starting point, a mean-field treatment would lead to a 
phase which is always unstable against phase separation, similar to the spiral instability in 
2D case\cite{ss}.

Furthermore, a series of detailed features obtained in various numerical work are also 
reproduced at this mean-field level. The energy spectra of magnons and quasiparticles 
have been determined in the $C1S0$ phase where they all exhibit finite gaps varying with 
the doping concentration. The gap of the former continuously evolves from the insulating 
phase while the one of the latter arises from the formation of Cooper pairs between 
quasiparticles. The minimum gap of creating a pair of quasiparticle excitations is smaller 
than that of the magnon. This indicates that the spin gap in the two-leg ladder system is 
generally associated with quasiparticles and shows a discontinuous evolution with doping, 
as first pointed out by Troyer {\it et al.}\cite{Troyer}. We also examine the local structure 
of hole pairs and show that the pairing on diagonal sites occurs simultaneously with the 
condensation of rung hole pairs. This point was also noted previously by 
Sierra {\it et al.}\cite{Sierra97}. Since the pairing between quasiparticles also results from 
the condensation of rung hole pairs, a relationship between the pairing on diagonal sites and 
the spin gap in the two-leg ladders is then established. 
        
The rest of the paper is organized as follows: In Section 2, we introduce the phase string 
formulation and bond-operator representation. The general feature of the resulting 
Hamiltonian is discussed and the mean-field treatment is presented in Section 3. In Section 
4, we present our numerical analysis of the mean-field equations. The finial section is 
devoted to a conclusive discussion. For the sake of compactness, some useful and relevant 
formulae are listed in the Appendices. 
 
%%%%%%%%%%%%%%%%%%%%%%%%%%%%%%%%%%%%%%%%%%%%

\section{Mathematical formulation}

We start with the original $t-J $ Hamiltonian  
\begin{eqnarray}
{\rm H}_{t-J}=P_s\{-t\sum_{\langle i,j\rangle }(c_{i\sigma }^{+}c_{j\sigma
}+H.c.)+J\sum_{\langle i,j\rangle }({\bf S}_i\cdot {\bf S}_j-\frac 14
n_in_j)\}P_s
\end{eqnarray}
where $P_s$ is the projection operator which imposes the no-double-occupancy constraint 
such that the electron occupation number $n_i\leq 1$. The conventional way to handle the 
constraint is to introduce the so-called slave-particle representation of electron operator: 
$c_{j,\sigma }\rightarrow f_j^{+}b_{j,\sigma }$ such that the constraint $n_i\leq 1$ is 
replaced by an equality condition: $f^{\dagger}_if_i+\sum_{\sigma}b^{\dagger}_{i\sigma}
b_{i\sigma}=1$. For our purpose, in the following we will use the slave-fermion 
representation in which $f_j$ and $b_{j,\sigma }$ satisfy the canonical anti-commutation
and commutation relations, respectively. Then the $t-J$ Hamiltonian $H_{t-J}=H_t+H_J$ 
can be rewritten as follows:
\begin{eqnarray}
{\rm H}_t &=&-t 
%TCIMACRO{\dsum }
%BeginExpansion
\mathop{\displaystyle \sum_{\langle i,j \rangle}}
%EndExpansion
\{{\rm f}_{i}^{{\rm \dagger }}{\rm f}_{j} (\sigma){\rm b}_{j\sigma}^{{\rm
\dagger }} {\rm b}_{i\sigma }+{\rm H.c.}\},
\nonumber \\
{\rm H}_J &=&-\frac {J}{2} 
%TCIMACRO{\dsum }
%BeginExpansion
\mathop{\displaystyle \sum_{\langle i,j \rangle} }
%EndExpansion
{\rm b}_{i\sigma }^{{\rm \dagger }} {\rm b}_{j-\sigma
}^{{\rm \dagger }}{\rm b}_{j-\sigma
^{\prime }} {\rm b}_{i\sigma ^{\prime }}.  \label{tJ0}
\end{eqnarray}

In obtaining the above expressions, a replacement was made: $b_{j,\sigma}\rightarrow 
(-\sigma )^jb_{j,\sigma }$ where $\sigma =1,-1$ for spin up and down, respectively. 
In this representation, the matrix element of ${\rm H_{J}}$ always remains 
negative-definite which is equivalent to say that the Marshall sign\cite{Marsh} has been 
built into the basis\cite{Weng}. But then the sign $\sigma$ appearing in ${\rm H_{t}}$ 
indicates that holes dislike the Marshall sign hidden in the spin background, and their 
motion generally creates Marshall-sign mismatches on their paths known as phase 
strings. Since ${\rm H_{J}}$ respects the Marshall sign rule at low energy, the phase
strings cannot be repaired through the spin flip processes. Such a phase-string-type doping 
effect has been argued\cite{Weng} to be the key to understanding the evolution of the 
ground state at finite doping. In the following, we first give a brief review of the phase 
string formulation developed in Ref.\onlinecite{Weng} to deal with this nonlocal singular 
phase effect.

%%%%%%%%%%%%%%%%%%%%%%%%%%%%%%%%%%%%%%%%%%

\subsection{Phase string formulation}

The basic idea underlying the phase string formulation is to ``gauge away'' the original 
singular source of the phase string effect shown in ${\rm H_{t}}$ [Eq. (\ref{tJ0})] such 
that the resulting form of the Hamiltonian becomes treatable in a perturbative scheme. 
According to Ref.\cite{Weng}, this procedure can be realized through a unitary 
transformation: 
\begin{eqnarray}
U\equiv \exp {\{-\frac i2\sum_{j\ne l}n_j^h\theta _j(l)(1-n_l^h-\sum_\sigma
\sigma n_{l,\sigma }^b)\}}  \label{U}
\end{eqnarray}
where $n_j^h$ and $n_{j,\sigma }^b$ are the number densities of holons and spinons with 
spin $\sigma $ at site $j$. Under the unitary transformation (\ref{U}), the electron operators 
become 
\begin{equation}
c_{j,\sigma }\rightarrow \tilde{h}_j^{+}\tilde{b}_{j,\sigma }(-\sigma
)^j\sigma ^{N_h}  \label{c}
\end{equation}
where 
\begin{eqnarray}
\tilde{h}_j^{+} &\equiv &h_j^{+}\exp {\{\frac i2\sum_{l\ne j}\theta
_j(l)(\sum_\sigma \sigma n_{l,\sigma }^b-1)\}},  \nonumber \\
\tilde{b}_{j,\sigma } &\equiv &b_{j,\sigma }\exp {\{-\frac i2\sigma
\sum_{l\ne j}\theta _j(l)n_l^h\}},  \nonumber \\
h_j &\equiv &f_j\exp {\{-i\sum_{l\ne j}\theta _j(l)n_l^h\}}.  \label{hb}
\end{eqnarray}
$N_h$ is the total number of holes. It is easy to verify that $h_j$ is a hard-core boson, i.e. 
they satisfy the following commutation relations: $[h_i,h_j]=0=[h_i,h_j^{+}]$, $i\neq j$ 
and $\{h_i,h_i\}=0$, $\{h_i,h_i^{+}\}=1 $. Even though in the original definition, 
$\theta_j(l)= \mbox{Im ln}(z_j-z_l)$, the choice of $\theta _j(l)$ is equivalent to a kind of
gauge fixing and we will explicitly write down our choice later.

With eqs. (\ref{c}) and (\ref{hb}), the $t-J$ Hamiltonian becomes 
\begin{eqnarray}
{\rm H}_t &=&-t 
%TCIMACRO{\dsum }
%BeginExpansion
\mathop{\displaystyle \sum_{\langle i,j \rangle} }
%EndExpansion
\{(e^{iA_{ij}^f}){\rm h}_{i}^{{\rm \dagger }}{\rm h}_{j} (e^{i\sigma
A_{ji}^h}){\rm b}_{j\sigma}^{{\rm \dagger }} {\rm b}_{i\sigma }+{\rm H.c.}\},
\nonumber \\
{\rm H}_J &=&-\frac {J}{2} 
%TCIMACRO{\dsum }
%BeginExpansion
\mathop{\displaystyle \sum_{\langle i,j \rangle} }
%EndExpansion
(e^{i\sigma A_{ij}^h}){\rm b}_{i\sigma }^{{\rm \dagger }} {\rm b}_{j-\sigma
}^{{\rm \dagger }}(e^{i\sigma ^{\prime } A_{ji}^h}){\rm b}_{j-\sigma
^{\prime }} {\rm b}_{i\sigma ^{\prime }}  \label{tJ}
\end{eqnarray}
where the gauge phases $A_{ij}^f$ and $A_{ij}^h$ are defined by 
\begin{eqnarray}
A_{ij}^f &=&\frac 12%
%TCIMACRO{\dsum }
%BeginExpansion
\mathop{\displaystyle \sum }
%EndExpansion
\limits_{l\neq i,j}[\theta _i(l)-\theta _j(l)]\left( \sum\limits_\sigma
\sigma n_{l\sigma }^b-1\right) ,  \nonumber \\
A_{ij}^h &=&\frac 12%
%TCIMACRO{\dsum }
%BeginExpansion
\mathop{\displaystyle \sum }
%EndExpansion
\limits_{l\neq i,j}[\theta _i(l)-\theta _j(l)]n_l^h.  \label{Ahf}
\end{eqnarray}
In the case of one chain, we can choose $\theta_{i}(l)$ such that all these gauge phases 
vanish\cite{Weng}. Thus, all important phases are absorbed into eq. (\ref{c})
in a form of phase shifts\cite{Weng}. This is the prefect case that the phase string effect 
is completely ``gauged away'' from the Hamiltonian. We shall see later that the situation 
slightly changes in the ladder case.

We now focus ourselves on the two-leg ladder and define $\theta_{j,m}(l,n)$ as follows: 
\begin{eqnarray}
\theta_{j,m}(l,n) &=& 0, \ j>l,  \nonumber \\
& & \pi, \ j<l,  \nonumber \\
\theta_{j,m}(j,n) &=& \ \frac{\pi}{2}, \ m=1, n=2,  \nonumber \\
& & -\frac{\pi}{2}, \ m=2, n=1  \label{th}
\end{eqnarray}
where $m$, $n\in \{1,2\}$ are labels of legs. $1$ and $2$ indicate the upper
and lower chains, respectively. This convention fixes the gauge phases as 
\begin{eqnarray}
A_{j,j+1}^{1h} &=& \frac{\pi}{4}( n_{j,2}^{h}+n_{j+1,2}^{h}),  \nonumber \\
A_{j,j+1}^{2h} &=& -\frac{\pi}{4}( n_{j,1}^{h}+n_{j+1,1}^{h}),  \nonumber \\
A_{j,j+1}^{1f} &=& -\frac{\pi}{2}(1- S_{j,2}^{z}-S_{j+1,2}^{z}),  \nonumber
\\
A_{j,j+1}^{2f} &=& \frac{\pi}{2}( 1- S_{j,1}^{z}-S_{j+1,1}^{z})  \label{gp}
\end{eqnarray}
where $n^{h}_{j,m}$ and $S^{z}_{j,m}$ are the number density of holes and
the $z$- component spin operator at site $j$ and chain $m$, respectively. 
$A_{j,j+1}^{mh(f)}$ represents the gauge phase on chain $m$. In the derivation of 
eq. (\ref{gp}), we have used two identities: $\exp{i\pi(1-2S^{z}_{j,m}+
\sigma n^{h}_{j,m})}=1$ and $\exp{i\pi (\sigma-\sigma^{\prime})n^{h}_{j,m}}=1$, 
which are valid in the physical Hilbert space. With the help of eq. (\ref{gp}), we can 
rewrite the $t-J$ Hamiltonian on two-leg ladders as follows: 
\begin{eqnarray}
{\rm H}_{t} &=& -t \sum_{j,\sigma}(h^{+}_{j,2}b_{j,2,\sigma}h_{j,1}
b^{+}_{j,1,\sigma}+H.c.)  \nonumber \\
& & -t \sum_{j,\sigma}\{ie^{-i\frac{\pi}{2}(S^{z}_{j,2}+S^{z}_{j+1,2})} e^{i%
\frac{\pi}{4}\sigma (n^{h}_{j,2}+n^{h}_{j+1,2})}h^{+}_{j+1,1}
b_{j+1,1,\sigma}h_{j,1}b^{+}_{j,1,\sigma}  \nonumber \\
& & -i e^{i\frac{\pi}{2}(S^{z}_{j,1}+S^{z}_{j+1,1})}e^{-i\frac{\pi}{4}
\sigma (n^{h}_{j,1}+n^{h}_{j+1,1})}h^{+}_{j+1,2}b_{j+1,2,\sigma}
h_{j,2}b^{+}_{j,2,\sigma}+H.c.\},  \nonumber \\
{\rm H}_{J} &=& -\frac{J_{\perp}}{2}\sum_{j}(\sum_{\sigma}b^{+}_{j,1,\sigma}
b^{+}_{j,2,-\sigma})(\sum_{\sigma^{\prime}}b_{j,2,-\sigma^{\prime}}
b_{j,1,\sigma^{\prime}})  \nonumber \\
& & -\frac{J}{2}\sum_{j,\sigma ,m}(b^{+}_{j,m,\sigma}b_{j,m,\sigma}
b^{+}_{j+1,m,-\sigma}b_{j+1,m,-\sigma})  \nonumber \\
& & -\frac{J}{2}\sum_{j,\sigma}\{e^{i\frac{\pi}{2}\sigma (n^{h}_{j,2}+
n^{h}_{j+1,2})}b^{+}_{j,1,\sigma}b_{j,1,-\sigma}
b^{+}_{j+1,1,-\sigma}b_{j+1,1,\sigma}  \nonumber \\
& & +e^{-i\frac{\pi}{2}\sigma(n^{h}_{j,1}+n^{h}_{j+1,1})}
b^{+}_{j,2,\sigma}b_{j,2,-\sigma}b^{+}_{j+1,2,-\sigma}b_{j+1,2,\sigma}\}.
\label{tJ2}
\end{eqnarray}
Note that in eq. ({\ref{tJ2}), the coupling between different legs is explicitly 
distinguished from the one on the same leg. In the following, we will reformulate 
it in the bond-operator representation under the implicit assumption that 
$J_{\perp}\gg t, J$.

     In contrast to the results of the single chain (see Ref.\cite{Weng}), there are 
some phase factors left in the Hamiltonian. If we increase the number of chains 
to infinity, they will turn out to become topological gauge fields. It is these gauge
fields that strongly affect the dynamics of holons and spinons in two 
dimensions\cite{Weng98}. On ladders, there are more than one path to connect 
two points while there is only one way to do it on the single chain. Consequently, 
in general, there is no closed path in the latter case\cite{foot4}. That is why we 
are unable to see those "gauge interactions" arising from the phase string explicitly 
present in the 1D $t-J$ model. As for the ladder case, there is no way to 
completely gauge away these phase factors no matter how we choose the 
$\theta_{j,m}(l,n)$. Nevertheless, those phase factors in eq. (\ref{tJ2}) will 
become trivial in the bond-operator representation introduced below. 

%%%%%%%%%%%%%%%%%%%%%%%%%%%%%%%%%%%%%%%%%
\subsection{Bond-operator description}

In half-filled case, the two-leg ladder $t-J$ model at $J_{\perp}\gg J$ 
can be well described\cite{Gopalan} 
at the mean-field level based on the bond-operator representation originally introduced 
by Sachdev and Bhatt\cite{Sachdev}. Such a description can be easily generalized to 
the doped case as emphasized in the Introduction, At each rung, the physical Hilbet 
space is spanned by nine states which can be generated by applying the bond operators 
to the vacuum state $\mid \phi_{0}\rangle$ as follows: 
\begin{eqnarray}
d^{+}_{j}\mid \phi_{0}\rangle &=& h^{+}_{j,1}h^{+}_{j,2}\mid 0 \rangle , 
\nonumber \\
a^{+}_{j,\sigma }\mid \phi_{0}\rangle &=&
(-\sigma)^{j+1}h^{+}_{j,1}b^{+}_{j,2,\sigma} \mid 0 \rangle , \ \ \bar{a}%
^{+}_{j,\sigma }\mid \phi_{0}\rangle = (-\sigma)^{j}b^{+}_{j,1,\sigma}
h^{+}_{j,2}\mid 0 \rangle ,  \nonumber \\
s^{+}_{j}\mid \phi_{0}\rangle &=& \frac{(-1)^{j}}{\sqrt{2}}(\mid \uparrow
\downarrow \rangle + \mid \downarrow \uparrow \rangle ),  \nonumber \\
t^{+}_{j,0}\mid \phi_{0}\rangle &=& \frac{(-1)^{j}}{\sqrt{2}}(\mid \uparrow
\downarrow \rangle - \mid \downarrow \uparrow \rangle ), \ \ t^{+}_{j,\sigma
}\mid \phi_{0}\rangle = \mid \sigma , \sigma \rangle .  \label{bond}
\end{eqnarray}
Here $\mid \sigma , \sigma^{\prime} \rangle \equiv b^{+}_{j,1,\sigma}
b^{+}_{j,2,\sigma^ {\prime}}\mid 0 \rangle$ and $\mid \phi_{0}\rangle$ is annihilated by 
these bond operators. $s$ represents the spin singlet and $t_{\alpha}$ with $\alpha=\pm 1, 0$ 
represent spin triplet excitations. $a_{\sigma}$ and $\bar{a}_{\sigma}$ particles carry the same 
quantum numbers as electrons. $d$ particles are spinless and charge two. We choose $s$ and $t$ 
operators to satisfy canonical commutation relations as in Ref.\cite{Gopalan} while $d$, 
$a_\sigma$, and $\bar{a}_\sigma$ operators are hard-core bosons because they contain $h$ 
operators. The prefactor, $(-1)^{j}$, in eq. (\ref{bond}) arises from $(-\sigma)^{j}$ in 
eq. (\ref{c}). The no-double-occupancy condition is replaced by the following one: 
\begin{eqnarray}
s^{+}_{j}s_{j}+\sum_{\alpha = \pm 1,
0}t^{+}_{j,\alpha}t_{j,\alpha}+\sum_{\sigma} (a^{+}_{j,\sigma}a_{j,\sigma}+%
\bar{a}^{+}_{j,\sigma}\bar{a}_{j,\sigma})+ d^{+}_{j}d_{j} &=& 1.
\label{cons}
\end{eqnarray}

By using eq. (\ref{bond}), one can express the bilinear operators composed of $h$ and 
$b_{\sigma}$ by these bond operators. Their detailed forms are left in the Appendix 
{\bf A}. We now obtain a Hamiltonian in which those gauge phases can be evaluated 
explicitly by substituting eq. (\ref{hbbb}) into eq. (\ref{tJ2}). The resulting Hamiltonian 
can be divided into three parts: 
\begin{eqnarray}
{\rm H}_{t-J} &=& {\rm H}_{0}+{\rm H}_{1}+{\rm H}_{2}  \nonumber
\end{eqnarray}
where 
\begin{eqnarray}
{\rm H}_{0} &=& -N(\frac{J_{\perp}}{4}+\frac{J}{2})+ t \sum_{j}\sigma
(a^{+}_{j,\sigma}\bar{a}_{j,\sigma}+H.c.)  \nonumber \\
& & -\frac{t}{2}\sum_{j}\{[s_{j+1}s^{+}_{j}+2d_{j+1}d^{+}_{j}+t_{j+1,0}t^{+}_{j,0}
        -2t_{j+1,\sigma}t^{+}_{j,\sigma}]  \nonumber  \\
& & \cdot (a^{+}_{j+1,\sigma}a_{j,\sigma}+\bar{a}^{+}_{j+1,\sigma} \bar{a}_{j,\sigma})
        +H.c.\}  \nonumber \\
& & +( \frac{J_{\perp}}{4}+\frac{J}{2})\sum_{j}(a^{+}_{j,\sigma}a_{j,\sigma}
+\bar{a}^{+}_{j,\sigma}\bar{a}_{j,\sigma})+( \frac{J_{\perp}}{4}+J)
\sum_{j}d^{+}_{j}d_{j}  \nonumber \\
& & -\frac{J}{4}\sum_{j}\{(a^{+}_{j,\sigma}a_{j,\sigma}+\bar{a}%
^{+}_{j,\sigma} \bar{a}_{j,\sigma})d^{+}_{j+1}d_{j+1}+(j \leftrightarrow
j+1)\}  \nonumber \\
& & - \mu \sum_{j}(2d^{+}_{j}d_{j}+a^{+}_{j,\sigma}a_{j,\sigma} +\bar{a}%
^{+}_{j,\sigma}\bar{a}_{j,\sigma})  \nonumber \\
& & +\frac{J}{2}\sum_{j}(s_{j}s^{+}_{j+1}t^{+}_{j,\alpha}t_{j+1,\alpha}
+s_{j}s_{j+1}t^{+}_{j,\alpha}t^{+}_{j+1,-\alpha}+H.c.)  \nonumber \\
& & +\frac{J_{\perp}}{4}\sum_{j}(-3s^{+}_{j}s_{j}+t^{+}_{j,\alpha}t_{j,%
\alpha})  \nonumber \\
& & -\sum_{j}\lambda_{j}(s^{+}_{j}s_{j}+t^{+}_{j,\alpha}t_{j,\alpha}+
a^{+}_{j,\sigma}a_{j,\sigma}+\bar{a}^{+}_{j,\sigma}\bar{a}_{j,\sigma}
+d^{+}_{j}d_{j}-1)  \label{H0}
\end{eqnarray}
and 
\begin{eqnarray}
{\rm H}_{1} &=& \frac{t}{\sqrt{2}}\sum_{j}\sigma
\{(s_{j+1}d_{j}+d_{j+1}s_{j}) (a^{+}_{j+1,\sigma}\bar{a}^{+}_{j,-\sigma}-%
\bar{a}^{+}_{j+1,\sigma} a^{+}_{j,-\sigma})+H.c.\}  \nonumber \\
& & -\frac{J}{2}\sum_{j}\{(\sum_{\sigma}a^{+}_{j+1,\sigma}a^{+}_{j,-\sigma})
(\sum_{\sigma^{\prime}}a_{j,-\sigma^{\prime}}a_{j+1,\sigma^{\prime}})+ (a
\rightarrow \bar{a})\}  \nonumber \\
& & -\frac{J}{2}\sum_{j}d^{+}_{j}d_{j}d^{+}_{j+1}d_{j+1}.  \label{H1}
\end{eqnarray}
Here $\mu$ is the chemical potential of holes. $\lambda_{j}$ is the Lagrangian multiplier to 
impose the constraint (\ref{cons}). $N$ is the number of sites for the single chain. The 
attraction between $d$ particles comes from the $-1/4n_{i}n_{j}$ term while those among 
quasiparticles arise from the exchange term. All the remaining terms are collected in 
${\rm H}_{2}$, which consists of terms that involve either spin-flip processes or the creation 
or annihilation of triplet excitations. We shall see later that all spin excitations are gapped. 
Thus, we expect that the inclusion of ${\rm H}_{2}$ is supposed not to change the main 
features of our results and will not consider it in the following calculations. We leave the 
detailed form of ${\rm H}_{2}$ in the Appendix {\bf B} and define our working Hamiltonian 
as ${\rm H}\equiv {\rm H}_{0}+{\rm H}_{1}$ in the following sections.

%%%%%%%%%%%%%%%%%%%%%%%%%%%%%%%%%%%%%%%%%%%%%%

\section{Mean-Field Theory}

\subsection{The low energy effective Hamiltonian}

Before taking the mean-field approximation, we would like to re-organize the general form of 
${\rm H_{t-J}}$ in the phase-string and bond-operator representation 
[Eqs. (\ref{H0}) and (\ref{H1})]. Then some features of it can be more easily revealed. 

We first express the $a$-operators in terms of the bonding and anti-bonding operators as follows: 
\begin{equation}
a_{j,\sigma} = \frac{1}{\sqrt{2}}(a_{-,j,\sigma}+i\sigma a_{+,j,\sigma}), \ 
\bar{a}_{j,\sigma} = \frac{1}{\sqrt{2}}(\sigma
a_{-,j,\sigma}-ia_{+,j,\sigma})  \label{ee}
\end{equation}
where $a_{\pm ,\sigma}$ denote the bonding and anti-bonding operators, respectively. Since 
$a_{\pm ,\sigma}$ operators are still hard-core bosons, one may introduce the following 
Jordan-Wigner transformation to transform them into fermions without changing the 
Hamiltonian:
\begin{equation}
a^{+}_{+,j,\sigma} = e^{+}_{j,\sigma}U_{j}, \ a^{+}_{-,j,\sigma} = \bar{e}%
^{+}_{j,\sigma}U_{j}.  \label{ee2}
\end{equation} 
Here $U_{j}= \exp{\{i \pi \sum_{l < j,\sigma}(a^{+}_{+,l,\sigma}a_{+,l,\sigma}
+a^{+}_{-,l,\sigma}a_{-,l,\sigma})\}}$. $e_{j,\sigma}$ and $\bar{e}_{j,\sigma}$ become 
fermions and satisfy the canonical anti-commutation relations.

Secondly, because under the unitary transformation (\ref{U}), $S^{+}_{j}$ becomes 
\begin{eqnarray}
S^{+}_{j,1} &=& (-1)^{j}b^{+}_{j,1,\uparrow}b_{j,1,\downarrow}\exp{\{i\pi
\sum_{l < j} (n^{h}_{l,1}+n^{h}_{l,2})-i\frac{\pi}{2}n^{h}_{j,2}\}}, 
\nonumber \\
S^{+}_{j,2} &=& (-1)^{j+1}b^{+}_{j,2,\uparrow}b_{j,2,\downarrow}\exp{\{i\pi
\sum_{l < j} (n^{h}_{l,1}+n^{h}_{l,2})+i\frac{\pi}{2}n^{h}_{j,1}\}}
\label{SS}
\end{eqnarray}
instead of $(-1)^{j}b^{+}_{j,\uparrow}b_{j,\downarrow}$, the $t$ operators defined based on 
$b^{+}_{j,\sigma}$ in (\ref{bond}) do not form a vector. Therefore, the coefficients before the 
terms $t_{j+1,0}t^{+}_{j,0}$ and $t_{j+1,\sigma}t^{+}_{j,\sigma}$ in eq. (\ref{H0}) have 
different signs. The manifest rotational symmetry can be easily recovered by introducing the 
following unitary transformation: 
\begin{equation}
t^{+}_{j,\sigma} \ \rightarrow \ t^{+}_{j,\sigma}\exp{\{-i\pi \sigma \sum_{l
< j}(n^{h}_{l,1}+ n^{h}_{l,2})\}}.  \label{tsig}
\end{equation}
for $t_{j,\sigma}$ with $\sigma =\pm 1$.

Then ${\rm H_{t-J}}$ is simplified after substituting eqs. (\ref{ee}) and 
(\ref{ee2}) into eqs. (\ref{H0}) 
and (\ref{H1}) and performing the transformation (\ref{tsig})\cite{foot1}: 
\begin{eqnarray}
{\rm H}_{0} &=& -N(\frac{J_{\perp}}{4}+\frac{J}{2})-t \sum_{j}
(e^{+}_{j,\sigma} e_{j,\sigma}-\bar{e}^{+}_{j,\sigma}\bar{e}_{j,\sigma}) 
\nonumber \\
& & -\frac{t}{2} \sum_{j}\{[s_{j+1}s^{+}_{j}+2d_{j+1}d^{+}_{j}+t_{j+1,0}t^{+}_{j,0}
        +2t_{j+1,\sigma}t^{+}_{j,\sigma}]  \nonumber  \\
& & \cdot (e^{+}_{j+1,\sigma}e_{j,\sigma}+\bar{e}^{+}_{j+1,\sigma}\bar{e}_{j,\sigma})
        +H.c.\}  \nonumber \\
& & +(\frac{J_{\perp}}{4}+\frac{J}{2})\sum_{j}(e^{+}_{j,\sigma}e_{j,\sigma}+ 
\bar{e}^{+}_{j,\sigma}\bar{e}_{j,\sigma})+(\frac{J_{\perp}}{4}+J)\sum_{j}
d^{+}_{j}d_{j}  \nonumber \\
& & -\frac{J}{4}\sum_{j}\{(e^{+}_{j,\sigma}e_{j,\sigma}+\bar{e}%
^{+}_{j,\sigma} \bar{e}_{j,\sigma})d^{+}_{j+1}d_{j+1}+(j \leftrightarrow
j+1)\}  \nonumber \\
& & - \mu \sum_{j}(2d^{+}_{j}d_{j}+e^{+}_{j,\sigma}e_{j,\sigma} +\bar{e}%
^{+}_{j,\sigma}\bar{e}_{j,\sigma})  \nonumber \\
& & +\frac{J}{2}\sum_{j}(s_{j}s^{+}_{j+1}t^{+}_{j,\alpha}t_{j+1,\alpha}
+s_{j}s_{j+1}t^{+}_{j,\alpha}t^{+}_{j+1,-\alpha}+H.c.)  \nonumber \\
& & +\frac{J_{\perp}}{4}\sum_{j}(-3s^{+}_{j}s_{j}+t^{+}_{j,\alpha}t_{j,%
\alpha})  \nonumber \\
& & - \sum_{j}\lambda_{j}(s^{+}_{j}s_{j}+t^{+}_{j,\alpha}t_{j,\alpha}+
e^{+}_{j,\sigma}e_{j,\sigma}+\bar{e}^{+}_{j,\sigma}\bar{e}_{j,\sigma}+
d^{+}_{j}d_{j}-1)  \label{H0e}
\end{eqnarray}
and 
\begin{eqnarray}
{\rm H}_{1} &=& -\frac{t}{\sqrt{2}}\sum_{j} \{(s_{j+1}d_{j}+d_{j+1}s_{j})
(e^{+}_{j+1,\sigma}e^{+}_{j,-\sigma}-\bar{e}^{+}_{j+1,\sigma} \bar{e}%
^{+}_{j,-\sigma})+H.c.\}  \nonumber \\
& & -\frac{J}{4}\sum_{j,\sigma ,\sigma^{\prime}}\{(e^{+}_{j+1,\sigma}
e^{+}_{j,-\sigma}+\bar{e}^{+}_{j+1,\sigma}\bar{e}^{+}_{j,-\sigma})
(e_{j,-\sigma^{\prime}}e_{j+1,\sigma^{\prime}}+\bar{e}_{j,-\sigma^{\prime}} 
\bar{e}_{j+1,\sigma^{\prime}})  \nonumber \\
& & +(e^{+}_{j+1,\sigma}\bar{e}^{+}_{j,-\sigma}-\bar{e}^{+}_{j+1,\sigma}
e^{+}_{j,-\sigma})(e_{j,-\sigma^{\prime}}\bar{e}_{j+1,\sigma^{\prime}}- \bar{%
e}_{j,-\sigma^{\prime}}e_{j+1,\sigma^{\prime}})\}  \nonumber \\
& & -\frac{J}{2}\sum_{j}d^{+}_{j}d_{j}d^{+}_{j+1}d_{j+1}.  \label{H1e}
\end{eqnarray}

In addition to the spin rotational symmetry, translation symmetry, and the electromagnetic U(1) 
symmetry, we note that the Hamiltonian of the two-leg ladder is also invariant under the 
exchange of chain indices to which we can assign a parity operator. Under the exchange of 
chain indices, the bond operators transform as follows: $s \rightarrow s, \ t_{\alpha} \rightarrow 
-t_{\alpha}, \ d \rightarrow d$. For quasiparticles, it is $a_{\sigma} \leftrightarrow -\sigma 
\bar{a}_{\sigma}$, or $e_{\sigma} \rightarrow e_{\sigma}, \ \bar{e}_{\sigma} \rightarrow 
-\bar{e}_{\sigma}$. We see that magnons and quasiparticles in the anti-bonding band are parity 
odd while $d$ bosons and quasiparticles in the bonding band are parity even\cite{foot2}.

We shall see later that both quasiparticles and magnons have gaps at low doping concentration. 
Consequently, we can integrate them out (treating $s$ as a c-number.) and obtain the low 
energy theory described by the following hard-core boson (HCB) model: 
\begin{equation}
{\rm H} = -t^{*}\sum_{j}(d^{+}_{j+1}d_{j}+H.c.)+V\sum_{j}
d^{+}_{j}d_{j}d^{+}_{j+1}d_{j+1}
\label{effH}
\end{equation}
where $t^{*}$ is the effective hopping amplitude and the dominant contribution to $V$ comes 
from the attraction between $d$ particles in eq.(\ref{H1e}). This model can be solved by a 
bosonization approach\cite{Haldane}. For $V < -2\mid t^{*} \mid$ the system is phase 
separated and this occurs only at very large values of $J_\perp$ for physically reasonable values 
of $J/t$. For example, $J_\perp > 31.8t$ for $J/t=0.5$ (see Troyer {\it et al.}\cite{Troyer}). In 
the region where the system is stable against the phase separation, the low-lying excitation 
will be the phase fluctuations of $d$-particles, which corresponds to the collective charge mode.  

Although the HCB model, eq. (\ref{effH}), is appropriate to describe the low energy properties 
of two-leg ladders in the lightly doped region, it can not address questions such as the internal 
structure of hole pairs, which is related to the nature of the superconducting order parameter, 
and how the spectra of those gapped modes vary with the hole concentration. Later we will 
answer these questions by a mean-field treatment of eqs. (\ref{H0e}) and (\ref{H1e}).   

%%%%%%%%%%%%%%%%%%%%%%%%%%%%%%%%%%%%%%%%%%%%%%

\subsection{Mean-field equations}

To proceed with the mean-field approximation, we first note that the undoped two-leg ladder is 
characterized as a spin liquid with non-vanishing RVB order parameter $\langle s_{j} \rangle$. 
Following Gopalan {\it et al.}\cite{Gopalan}, we take the ansatz for the spin part as: 
$\langle s_{j} \rangle = \bar{s}, \ \langle t_{j,\alpha} \rangle = 0, 
\ Q_{\alpha}= \langle t_{j+1,\alpha}t^{+}_{j,\alpha} \rangle$ , and 
$P_{\alpha}= \langle t^{+}_{j,\alpha}t_{j,\alpha} \rangle$ where $\alpha =\pm 1, 0$. Because 
of the rotational symmetry, $Q_{+}=Q_{-}=Q_{0}$ and $P_{+}=P_{-}=P_{0}$. We set 
$P \equiv P_{+}+P_{-}+P_{0}$ and $Q \equiv Q_{+}+Q_{-}+Q_{0}$. 

For the charge part at finite doping, we define the following mean-field parameters: 
$\chi_{\sigma}= \langle e^{+}_{j+1,\sigma}e_{j,\sigma}\rangle$ and 
$\bar{\chi}_{\sigma} = \langle \bar{e}^{+}_{j+1,\sigma}\bar{e}_{j,\sigma} \rangle$. Again 
due to the rotational symmetry, $\chi_{+}=\chi_{-}$ and $\bar{\chi}_{+}=\bar{\chi}_{-}$. 
We define $\chi = \chi_{+}+\chi_{-}$ and $\bar{\chi}= \bar{\chi}_{+}+\bar{\chi}_{-}$. 
Moreover, we take $\lambda_{j}=\lambda$ in accordance with the translational invariance 
along the chain direction.

Note that there is a linear $d$-operator appearing in the first term of ${\rm H}_{1}$ 
[Eq. (\ref{H1e})], describing the process that a rung hole pair dissolves into two quasiparticles 
or a pair of quasiparticles is recombined into a rung hole pair. Such a term (with $s_{j} 
\rightarrow \bar{s}$) looks similar to a Hubbard-Stratonovich transformed four-fermion 
interaction in the BCS theory with $d$ playing the role of the order parameter. Thus, we will 
treat $d_{j}$ as a c-number by assuming $d_{j}=\bar{d}$ and neglect its phase fluctuations. 
A solution with non-vanishing $\bar{d}$ will then immediately lead to forming Cooper pairs 
for $e_{\uparrow}$ and $e_\downarrow$ according to the first term in ${\rm H}_{1}$. To 
make things more transparent, let us define the following pairing fields for quasiparticles: 
\begin{equation}
\Delta \equiv \sum_{\sigma}\langle
e^{+}_{j+1,\sigma}e^{+}_{j,-\sigma}\rangle, \ \bar{\Delta} \equiv
\sum_{\sigma}\langle \bar{e}^{+}_{j+1,\sigma}\bar{e}^{+}_{j,-\sigma} \rangle.
\label{ds}
\end{equation}
Then 
\begin{eqnarray}
\Delta+\bar{\Delta} &=& -\sum_{\sigma} \langle
a^{+}_{j+1,\sigma}a^{+}_{j,-\sigma}- \bar{a}^{+}_{j+1,\sigma}\bar{a}%
^{+}_{j,-\sigma}\rangle ,  \nonumber \\
\Delta-\bar{\Delta} &=& -\sum_{\sigma}\sigma \langle a^{+}_{j+1,\sigma}\bar{a%
}^{+}_{j,-\sigma} -\bar{a}^{+}_{j+1,\sigma}a^{+}_{j,-\sigma}\rangle . 
\nonumber
\end{eqnarray}
Here $\Delta+\bar{\Delta}$ and $\Delta-\bar{\Delta}$ represent the hole pairing along the chain 
direction and diagonal sites, respectively and are shown schematically in Fig. 1. Now it is easy to 
see that the pairing on the rung and diagonal sites must occur simultaneously. This is what the 
first term of ${\rm H_{1}}$ tells us and it is consistent with the picture emerged from numerical 
studies\cite{White,Sierra97}. 

Naively, it seems that $\Delta$ and $\bar{\Delta}$ are not tensors under the spin rotation and 
thus our mean-field ansatz may break the rotational symmetry. This is, in fact, disguised by the 
phase string effect as we have discussed before. Our mean-field ansatz indeed respects 
rotational symmetry which is left to be discussed in the Appendix {\bf C}.

Based on the above mean-field ansatz, we finally obtain the following mean-field Hamiltonian: 
\begin{eqnarray}
{\rm H}_{MF} &=& N\{\lambda -\frac{J_{\perp}}{4}-\frac{J}{2}-(\frac{3}{4}%
J_{\perp}+ \lambda)\bar{s}^{2}+(\chi+\bar{\chi}) Qt+ \frac{J}{8}(\Delta+\bar{%
\Delta})^{2}  \nonumber \\
& & -\frac{J}{16}(\chi -\bar{\chi})^{2}-(\frac{J}{2}\bar{d}^{2}+2\mu
+\lambda -\frac{J_{\perp}}{4}-J)\bar{d}^{2}\}  \nonumber \\
& & -(t(\bar{d}^{2}+\frac{\bar{s}^{2}+Q}{2})-\frac{J}{16}(\chi -\bar{\chi}))
\sum_{j}(e^{+}_{j+1,\sigma}e_{j,\sigma}+H.c.)  \nonumber \\
& & -(t(\bar{d}^{2}+\frac{\bar{s}^{2}+Q}{2})+\frac{J}{16}(\chi -\bar{\chi}))
\sum_{j}(\bar{e}^{+}_{j+1,\sigma}\bar{e}_{j,\sigma}+H.c.)  \nonumber \\
& & +(-t-\mu -\lambda +\frac{J_{\perp}}{4}+\frac{J}{2}-\frac{J}{2}\bar{d}%
^{2}) \sum_{j}e^{+}_{j,\sigma}e_{j,\sigma}  \nonumber \\
& & +(t-\mu -\lambda +\frac{J_{\perp}}{4}+\frac{J}{2}-\frac{J}{2}\bar{d}%
^{2}) \sum_{j}\bar{e}^{+}_{j,\sigma}\bar{e}_{j,\sigma}  \nonumber \\
& & -(\frac{J}{8}(\Delta+\bar{\Delta})+\sqrt{2}t\bar{s}\bar{d})\sum_{j}
(e^{+}_{j+1,\sigma}e^{+}_{j,-\sigma}+H.c.)  \nonumber \\
& & -(\frac{J}{8}(\Delta+\bar{\Delta})-\sqrt{2}t\bar{s}\bar{d})\sum_{j} (%
\bar{e}^{+}_{j+1,\sigma}\bar{e}^{+}_{j,-\sigma}+H.c.)  \nonumber \\
& & +\frac{1}{2}\sum_{j}\{(J\bar{s}^{2}-t(\chi+\bar{\chi}))t^{+}_{j,\alpha}
t_{j+1,\alpha}+J\bar{s}^{2}t^{+}_{j,\alpha}t^{+}_{j+1,-\alpha}+H.c.\} 
\nonumber \\
& & +(\frac{J_{\perp}}{4}-\lambda)\sum_{j} t^{+}_{j,\alpha}t_{j,\alpha}.
\label{HMF}
\end{eqnarray}
Eq. (\ref{HMF}) can be diagonalized by Bogolioubov transformations. 
We leave the procedure in the Appendix {\bf D} and write down the
diagonalized Hamiltonian in the following:
\begin{eqnarray}
{\rm H}_{MF} &=& \Omega_{0}+\sum_{k}\{E_{k}
(\alpha^{+}_{k}\alpha_{k}+\beta^{+}_{k} \beta_{k})+\bar{E}_{k} (\bar{\alpha}%
^{+}_{k}\bar{\alpha}_{k}+ \bar{\beta}^{+}_{k}\bar{\beta}_{k})\}  \nonumber \\
& & +\sum_{k}\omega_{k}\gamma^{+}_{k\alpha}\gamma_{k\alpha}  \label{HMF3}
\end{eqnarray}
where 
\begin{eqnarray}
\Omega_{0} &=& \sum_{k}(\frac{3}{2}\omega_{k}-E_{k}-\bar{E}_{k})+N\{\frac{\lambda%
}{2} -2\mu -\frac{J_{\perp}}{8}+\frac{J}{2}+(\chi+\bar{\chi}) Qt- (\frac{3}{4%
}J_{\perp}+\lambda)\bar{s}^{2}  \nonumber \\
& & +\frac{J}{8}(\Delta+\bar{\Delta})^{2}-\frac{J}{16}(\chi -\bar{\chi}%
)^{2}- (\frac{J}{2}\bar{d}^{2}+2\mu +\lambda -\frac{J_{\perp}}{4})\bar{d}%
^{2}\}  \label{E0}
\end{eqnarray}
is the mean-field ground state energy. (In fact, $\Omega_{0}$ is the zero
temperature grand potential.)

The parameters $\bar{d}$, $\bar{s}$, $\mu$, $\lambda$, $\chi$, $\bar{\chi}$, 
$\Delta$, $\bar{\Delta}$, $P$, and $Q$ are determined by solving eq. (\ref
{ds}), the following self-consistent equations: 
\begin{eqnarray}
Q &=& \sum_{\alpha}\langle t_{j+1\alpha}t^{+}_{j\alpha}\rangle, \ P =
\sum_{\alpha}\langle t^{+}_{j\alpha}t_{j\alpha}\rangle,  \nonumber \\
\chi &=& \sum_{\sigma}\langle e^{+}_{j+1\sigma}e_{j\sigma}\rangle, \ \bar{%
\chi} = \sum_{\sigma}\langle \bar{e}^{+}_{j+1\sigma}\bar{e}_{j\sigma}\rangle,
\label{selfeq}
\end{eqnarray}
and the saddle-point equations: 
\begin{eqnarray}
\frac{\partial \Omega_{0}}{\partial \lambda} &=& 0, \ \ \frac{\partial \Omega_{0}}
{\partial \bar{s}} \ = \ 0,  \nonumber \\
\frac{\partial \Omega_{0}}{\partial \bar{d}} &=& 0, \ \ \frac{\partial \Omega_{0}}
{\partial \mu} \ = \ -2N\delta,  \label{variation}
\end{eqnarray}
where $\delta$ is the hole concentration. We obtain the following mean-field equations
from eqs. (\ref{ds}), (\ref{selfeq}), and (\ref{variation}): 
\begin{eqnarray}
1-\bar{s}^{2}+\bar{d}^{2} &=& P+2\delta,  \label{mfe1} \\
1+\bar{d}^{2}-\delta &=& \frac{1}{2}\frac{1}{N}\sum_{k}(\frac{\epsilon_{k}}{%
E_{k}}+ \frac{\bar{\epsilon}_{k}}{\bar{E}_{k}}),  \label{mfe2}
\end{eqnarray}
\begin{eqnarray}
(2\frac{t}{J}(\chi+\bar{\chi})+2\frac{\mu}{J} +\frac{\lambda}{J} - \frac{%
J_{\perp}}{4J}-1 +\delta )\bar{d} &=& -\sqrt{2}\frac{t}{J}\bar{s}(\Delta-%
\bar{\Delta}),  \label{mfe3} \\
\frac{3J_{\perp}}{4J}+\frac{\lambda}{J}-Q+\frac{t}{J}(\chi+\bar{\chi}) +%
\sqrt{2}\frac{t\bar{d}}{J\bar{s}}(\Delta-\bar{\Delta}) &=& -3\frac{1}{N}%
\sum_{k}\cos{k}\frac{\Pi_{k}}{\omega_{k}},  \label{mfe4}
\end{eqnarray}
\begin{eqnarray}
\Delta &=& \frac{1}{N}\sum_{k}\sin{k}\frac{\Gamma_{k}}{E_{k}}, \ \bar{\Delta}
= \frac{1}{N}\sum_{k}\sin{k}\frac{\bar{\Gamma}_{k}}{\bar{E}_{k}},
\label{mfe5} \\
\chi &=& -\frac{1}{N}\sum_{k}\cos{k}\frac{\epsilon_{k}}{E_{k}}, \ \bar{\chi}
= -\frac{1}{N}\sum_{k}\cos{k}\frac{\bar{\epsilon}_{k}}{\bar{E}_{k}},
\label{mfe6} \\
Q &=& \frac{3}{2}\frac{1}{N}\sum_{k}\cos{k}\frac{\Lambda_{k}}{\omega_{k}}, \
P = \frac{3}{2}\frac{1}{N}\sum_{k}\frac{\Lambda_{k}}{\omega_{k}}-\frac{3}{2}.
\label{mfe7}
\end{eqnarray}
Notice that eq. (\ref{mfe3}) says that $\Delta-\bar{\Delta}=0$ as long as $\bar{d}=0$. 

We close this section by a remark. When $\delta=0$, i.e., the undoped case,
our mean-field equations are not exactly reduced to those in Ref.\cite{Gopalan}. 
The difference arises from the fact that there are three components for $t$ 
operators and each contributes $\sum_{k}\frac{1}{2}\omega_{k}$ to $\Omega_{0}$. 
Thus, the zero-point energy is $\sum_{k}\frac{3}{2}\omega_{k}$ instead of 
$\sum_{k}\frac{1}{2}\omega_{k}$. A similar effect also appears in eq. (\ref{E0}) in 
which the coefficient of $N\lambda$ changes from $3/2$ to $1/2$. This point was 
neglected in Ref.\cite{Gopalan}. We will see later that this difference increases the spin 
gap in comparison with that obtained in Ref.\cite{Gopalan}.

%%%%%%%%%%%%%%%%%%%%%%%%%%%%%%%%%%%%%%%%%

\section{Results}

\subsection{Undoped case}

As a reference point, we first examine the results of our mean-field equations for the 
undoped case. By defining $\tilde{\Lambda}_{k}=\Lambda_{k}\mid_{\chi =0}$, the 
relevant integrals in the mean-field equations can be expressed as elliptic 
integrals\cite{Integral}: 
\begin{eqnarray}
\frac{1}{N}\sum_{k}\frac{\tilde{\Lambda}_{k}}{\omega_{k}} &=& \frac{1}{\pi}\{%
\frac{1} {\sqrt{1+\nu}}K(\sqrt{\frac{2\nu}{1+\nu}})+\sqrt{1+\nu}E(\sqrt{%
\frac{2\nu}{1+\nu}})\},  \nonumber \\
\frac{1}{N}\sum_{k}\cos{k}\frac{\tilde{\Lambda}_{k}-2\Pi_{k}}{\omega_{k}}
&=& -\frac{2} {\pi \nu}\{\frac{1}{\sqrt{1+\nu}}K(\sqrt{\frac{2\nu}{1+\nu}}) 
\nonumber \\
& & -\sqrt{1+\nu}E(\sqrt{\frac{2\nu}{1+\nu}})\}  \label{ellip}
\end{eqnarray}
where $K(\xi)$ and $E(\xi)$ are respectively the complete elliptic integrals of the first 
and second kind with modulus $\xi$. The dimensionless parameter $\nu$ is defined in the
Appendix {\bf D}.

Then, in the undoped limit, eqs. (\ref{mfe1}), (\ref{mfe2}), (\ref{mfe3}), (%
\ref{mfe4}), (\ref{mfe5}), (\ref{mfe6}), and (\ref{mfe7}) are reduced to the
following forms: 
\begin{eqnarray}
\frac{5}{2}-\bar{s}^{2} &=& \frac{3}{2\pi}\{\frac{1}{\sqrt{1+\nu}} K(\sqrt{%
\frac{2\nu}{1+\nu}})  \nonumber \\
& & +\sqrt{1+\nu}E(\sqrt{\frac{2\nu}{1+\nu}})\},  \label{undo1} \\
\frac{3}{4}+\frac{\lambda}{J_{\perp}} &=& -\frac{3\eta}{\pi \nu}\{\frac{1}{%
\sqrt{1+\nu}} K(\sqrt{\frac{2\nu}{1+\nu}})  \nonumber \\
& & -\sqrt{1+\nu}E(\sqrt{\frac{2\nu}{1+\nu}})\}  \label{undo2}
\end{eqnarray}
where $\eta = J/J_{\perp}$. The spin-triplet excitation spectrum is given by 
\begin{equation}
\omega_{k} = J_{\perp}(\frac{1}{4}-\frac{\lambda}{J_{\perp}})\sqrt{1+\nu \cos%
{k}}.  \label{tdisp}
\end{equation}
In eq. (\ref{tdisp}), we have to assume $0 \leq \nu \leq 1$; otherwise the mean-field 
equations would break down. This is verified in the following calculations. The band 
minimum is at $k=\pi$ and the spin gap is determined by 
\begin{equation}
\Delta_{t} = J_{\perp}(\frac{1}{4}-\frac{\lambda}{J_{\perp}})\sqrt{1-\nu}.
\end{equation}

We can analytically study the asymptotic behavior of the spin gap for small values of 
$\eta$ (and hence $\nu$) to see the difference between our equations and those in 
Ref.\cite{Gopalan}. For small values of $\xi$ the elliptic integrals $K(\xi)$ and 
$E(\xi)$ can be expanded in a power series as 
\begin{eqnarray}
K(\xi) &=& \frac{\pi}{2}(1+\frac{1}{4}\xi^{2}+\frac{9}{64}\xi^{4}- \cdots), 
\nonumber \\
E(\xi) &=& \frac{\pi}{2}(1-\frac{1}{4}\xi^{2}-\frac{3}{64}\xi^{4}- \cdots), 
\nonumber
\end{eqnarray}
and we obtain 
\begin{eqnarray}
\nu &=& 2\eta (1+\frac{23}{8}\eta^{2}+O(\eta^{4})),  \nonumber \\
\frac{1}{4}-\frac{\lambda}{J_{\perp}} &=& 1+\frac{3}{4}\eta^{2}+O(\eta^{4}),
\nonumber
\end{eqnarray}
and the spin gap is given by 
\begin{equation}
\Delta_{t} = J_{\perp}(1-\eta+\frac{1}{4}\eta^{2}+O(\eta^{3})).  \label{tgap}
\end{equation}
We find that our result, eq. (\ref{tgap}), is larger than that in Ref.\cite{Gopalan}, and 
is much closer to the strung rung interaction result, which is 
$J_{\perp}(1-\eta+\frac{1}{2}\eta^{2}+O(\eta^{3}))$\cite{Reig}. At the end of 
Sec. III we have pointed that there is a numerical factor missing in the zero-point 
energy in Ref.\cite{Gopalan} which is responsible for this discrepancy. 

To obtain $\Delta_{t}$ at any value of $J/J_{\perp}$, we numerically solved 
eqs. (\ref{undo1}) and (\ref{undo2}). The results are shown in Fig. 2. The spin gap at 
the instropic point, $\eta=1$, is about $0.501J$, which is very close to the numerical 
result --- $0.504J$\cite{Noack}. Of course, the mean-field approximation is only 
justified in the strong rung interaction regime and we do not expect the theory to be 
extended into the region with $\eta>1$ where the coupling between spins on the same 
leg becomes dominant over the rung coupling. In fact, our calculation shows that the 
spin gap continuously increases beyond $\eta > 1$, but according to Ref.\cite {Barnes}, 
it should smoothly diminish to zero as $\eta$ approaches $0$.

%%%%%%%%%%%%%%%%%%%%%%%%%%%%%%%%%%%%%%%%%%%%%

\subsection{Phases in doped case}
\subsubsection{The $C1S0$ phase}

The phase diagram obtained from numerical studies\cite{Poil,Troyer,Muller} shows that at 
most values of $J/t$ the two-leg ladders fall into the universality classes of Luther-Emery 
and Luttinger liquids for small and large doping concentration, respectively. The former and 
the latter are respectively denoted by $C1S0$ and $C1S1$ phases. ($CmSn$ means that 
there are $m$ gapless charge modes and $n$ gapless spin modes\cite{Fisher}.) 

The present theory using the bond-operator description presumably works at small doping 
for the Luther-Emery type phase. A mean-field solution with non-vanishing $\bar{d}$ is 
found at $\delta<0.5$ which is stable against the phase separation when $J/t$ is not too large.  
We interpret this mean-field state as the $C1S0$ phase. This is because in this situation there 
are gaps in the spectra of both quasiparticles and magnons. This can be understood as the 
following: To let quasiparticles be gapless, $\Gamma_k$ and $\bar{\Gamma}_k$ must be 
zero\cite{foot3}. From eq. (\ref{mfe5}), it results in vanishing $\Delta$ and $\bar{\Delta}$. 
This implies that $\bar{d}=0$ due to eq. (\ref{mfe3}). Therefore, a non-zero $\bar{d}$ will 
induce a gap for quasiparticles. The gap in the spectrum of magnons is a continuation of the 
one in the undoped case, which ensures the validity of the extension of the mean-field ansatz 
from the undoped case to the finite doping. The only gapless excitation in this region is the 
density fluctuations of hole pairs as discussed before.

{\bf Ground state energy:} To examine the validity of our mean-field ansatz, we first 
compute the ground state energy and compare it with numerical results. The mean-field 
ground state energy per site, $E_{0}$, is given as the following:
\begin{eqnarray}
 E_{0} &=& \frac{1}{2N}\{\Omega_{0}+\mu \sum_{j}(2\bar{d}^{2}+\langle 
                    a_{j\sigma}^{+}a_{j\sigma}+\bar{a}_{j\sigma}^{+}\bar{a}_{j\sigma}
                    \rangle )\}  \nonumber  \\
            &=& \frac{1}{2N}\Omega_{0}+\mu \delta .
\label{ener}
\end{eqnarray}
In the above derivation, we have used eq. (\ref{mfe2}). Also note that the number of sites
is $2N$. We calculated $E_{0}$ with $J/t=0.5$ and its doping dependence at various 
$J_{\perp}/J$'s is plotted in Fig. 3. By comparing with the numerical results in 
Ref.\cite{Sierra97}, we find that both the tendency of the energy versus the doping 
concentration and its absolute magnitude agree well with the data obtained by the recurrent 
variational ansatz (RVA) as well as the  density matrix renormalization group (DMRG) 
method. For comparison, in Fig. 3 the DMRG results\cite{Sierra97} at $\delta=1/8$ and 
$1/2$ are also shown. (Note that we set $J=1$ in Fig. 3 while $J=0.5$ in 
Ref.\cite{Sierra97} but $J/t=0.5$ is the same.) We see that the agreement is especially good 
in the region with large values of $J_{\perp}/J$ and small doping concentration as expected 
for the bond-operator representation. (The comparisons with the RVA results\cite{Sierra97} 
are even better over the whole $\delta \leq 0.5$ region.) Such a good agreement over a wide 
range of parameters indicates that our mean-field treatment based on the phase string and 
bond-operator formalism indeed captures the basic physics of doped two-leg ladders in the 
strong rung interaction regime. In the following, we focus on some detailed properties by 
solving the mean-field equations at $J/t=0.5$ and $J_{\perp}/J=10$.

{\bf Local structure of hole pairs:} Next, we would like to discuss the local structure of hole 
pairs in the $C1S0$ phase. As has been discussed in Ref.\cite{Sierra97}, holes will form 
pairs along the diagonal sites as well as along the rung and chain directions. This diagonal 
pairing is energetically favored by the $t$ term and the most probable configuration of two 
dynamical holes in a two-leg ladder\cite{White}. In our formalism, the amplitudes for 
pairing along the rung, diagonal, and chain directions can be respectively represented by 
$\bar{d}$, $\Delta-\bar{\Delta}$ and $\Delta+\bar{\Delta}$. We calculated 
$(\mid \Delta-\bar{\Delta}\mid)/\bar{d}$ and $(\mid \Delta+\bar{\Delta}\mid)/\bar{d}$ and 
plot them in Fig. 4. We found that both decrease with increasing $\delta$. In the low doping 
region, the amplitude of hole pairs on diagonal sites is almost comparable to the one of rung 
hole pairs and always larger than that on the chain direction. The amplitude of hole pairs 
along the leg being smaller than other hole configurations reflects the fact that the rung bonds 
are stronger than the leg bonds in the underlying two-leg spin ladder. The above results were 
also pointed out by Sierra {\it et al.}\cite{Sierra97}. They proposed a dimer hard-core boson 
(DHCB) model to describe the low energy properties of two-leg ladders in the strong rung 
interaction regime, which contains both the charge and spin degrees of freedom in contrast to 
the HCB model. The bond-operator formulation also retains these high energy modes. In 
particular, the coexistence of the diagonal pairing and rung hole pairs is further manifested in 
our approach.

{\bf Pairing symmetry:} We also calculated the expectation values of pairing fields along rung 
and chain directions. The corresponding operators are defined as follows: 
\begin{eqnarray}
\Delta_{x}(j) &\equiv& \frac{1}{\sqrt{2}}\sum_{\sigma}\sigma
c_{j+1,2,\sigma}c_{j,2,-\sigma},  \nonumber \\
\Delta_{y}(j) &\equiv& \frac{1}{\sqrt{2}}\sum_{\sigma}\sigma
c_{j,1,\sigma}c_{j,2,-\sigma}.  \label{pf}
\end{eqnarray}
These pairing fields can be represented by bond operators and we list their explicit forms in 
Appendix {\bf E}. In terms of the above mean-field parameters, their vacuum expectation 
values are given as follows: 
\begin{eqnarray}
i\langle \Delta_{x}\rangle &=& -\frac{1}{4\sqrt{2}}(\Delta +\bar{\Delta})[%
\bar{s}^{2}-2\bar{d}^{2} -\frac{3J_{\perp}}{4J}-\frac{\lambda}{J}  \nonumber
\\
& & +Q-\frac{t}{J}(\chi +\bar{\chi})-\sqrt{2}\frac{t\bar{d}}{J\bar{s}}
(\Delta -\bar{\Delta})],  \nonumber \\
i\langle \Delta_{y}\rangle &=& \frac{1}{2}\bar{d}\bar{s}.
\end{eqnarray}
The results are shown in Fig. 5. It is clear that there is a critical hole concentration 
$\delta_{c}$. (In our case, $\delta_{c}=0.37$ for $J/t=0.5$ and $J_{\perp}/J=10$.) In the 
low doping regime $\delta < \delta_{c}$, the pairing symmetry shows d-wave-like 
behavior while for $\delta > \delta_{c}$, it becomes s-wave-like symmetry. The difference 
can be attributed to different internal structures of hole pairs\cite{Sierra97}. When 
$\delta < \delta_{c}$, holes doped into a spin liquid state with RVB correlations form pairs 
with $d_{x^{2}-y^{2}}$-like structure. However, in the overdoped region, one moves into 
the low density limit characterized by electrons doped into a background with an internal 
s-wave-like symmetry. 

{\bf Spin excitations:} There are two kinds of excitations which carry non-trival spin 
quantum numbers. One is the magnon, which is represented by $t$ operators in our 
formulation and is the spin triplet excitation around ${\bf q}=(\pi , \pi)$. The other type of 
spin excitations are quasiparticles, which carry spin-$1/2$. The band minima of 
quasiparticles in bonding and anti-bonding bands are at ${\bf q}=(0,0)$ and 
${\bf q}=(0, \pi)$, respectively. The behaviors of their gaps varying with hole concentration 
are shown in Fig. 6. We found that the gap of magnons increases while quasiparticle gaps 
decrease with increasing $\delta$. In addition, it is easy to see that the low-lying spin modes 
with odd and even parity are magnons and quasiparticles in the bonding band, respectively. 
Real spin-$1$ excitations in two-leg ladders are composed of magnons or pairs of 
quasiparticles. The gap of the latter is still smaller than that of the former. Thus, the spin 
gap in the $C1S0$ phase is determined by quasiparticles instead of magnons. This was also 
shown by numerical studies\cite{Troyer}.

%%%%%%%%%%%%%%%%%%%%%%%%%%%%%%%%%%%%%%%%%%%

 \subsubsection{Phase separation} 

The above-discussed $C1S0$ superconducting phase may become unstable against phase 
separation when the value of $J/t$ becomes large in the $t-J$ ladders\cite{Troyer}. The 
reason is that in the large $J$ limit the gain in exchange energy by maximizing the number 
of AF bonds outweighs the cost in kinetic energy. 

The stability of the $C1S0$ solution against the phase separation can be examined by 
studying the compressibility $\kappa$:
\begin{eqnarray}
\kappa^{-1} &=& \delta^{2}\frac{\partial \mu}{\partial \delta}.  \nonumber
\end{eqnarray}
At $J_{\perp}/J=10$, we found that $\kappa$ diverges at $J/t=1.2$ and becomes negative 
when $J/t > 1.2$. This implies that our mean-field solutions with uniform hole density 
become unstable against phase separation when $J/t \geq 1.2$ [see Fig. 7(a)]. For 
$J_{\perp}/J=5$, the situation is similar as shown in Fig. 7(b). The only difference is that 
the boundary between the $C1S0$ phase and the phase separation region is moved to 
$J/t=1.6$. At $J_{\perp}/J=2$, the critical value of $J/t$ where the phase separation occurs 
not only increases but also becomes strongly doping-dependent as shown in Fig. 7(c). The 
latter trend is quite similar to that found in numerical studies\cite{Troyer,Sierra97} for the 
isotropic case, i.e. $J_\perp/J=1$. We note, however, that further reducing $J_{\perp}/J$ 
towards the isotropic limit in our mean-field theory does not improve more of the 
comparison with the numerical results since the ground-state energy starts to visibly deviate 
from numerical data at $J_{\perp}/J < 2$ even for small $\delta$ as shown in Fig. 3.

%%%%%%%%%%%%%%%%%%%%%%%%%%%%%%%%%%%%%%%%%%%%%%%%

\section{Discussions}

In this paper, we proposed a mean-field description on doped two-leg ladders in the strong 
rung interaction limit based on the phase string formulation. The transformation to bond 
operators is a natural choice in this formalism. With the help of bond operators, we can 
easily separate the high erergy and low energy processes in the $t-J$ Hamiltonian. Thus, a 
mean-field treatment becomes straightforward. Naively, there are two competing pairing 
channels in eq. (\ref{H1e}) - the pairing between $e_{\sigma}$ and $\bar{e}_{-\sigma}$ 
and the one between $e_{\sigma}$ and $e_{-\sigma}$ (or $\bar{e}_{\sigma}$ and 
$\bar{e}_{-\sigma}$ ). However, the condensation of the hard-core boson $d$ and a 
non-vanishing RVB order parameter $\bar{s}$ demand that the latter dominates. 
Furthermore, this type of pairing implies the formation of spin singlet bonds not only along 
the chain direction but also on diagonal sites. The latter turns out to be an important low 
energy structure in various types of $t-J$ models\cite{White}. This singlet bond becomes 
a strong nearest-neighbor one after one of the holes hops next to the other. As a consequence, 
the formation of this kind of singlets can maximize the hopping overalp with other hole 
configurations and lower its energy. This feature is a necessary result in our formulae as 
shown in eq. (\ref{mfe3}) or the linear $d$ term in ${\rm H_{1}}$ [Eq. (\ref{H1e})]. We 
have to emphasize that this mechanism for pairing comes from the $t$ term and is quite 
different from the "broken-bond" effect though the latter does enhance hole pairing somewhat. 
Also, the pairing {\it must} cause a gap opened up in the quasiparticle spectrum. Thus, all spin 
excitations are gapped in this region.

Here we would like to make some comments on the effects of phase string in two-leg ladders. 
If we directly apply the bond-operator representation to the original $t-J$ Hamiltonian without 
explicitly taking into account the nonlocal effect of phase string, then we would obtain a 
Hamiltonian with a different form in the $t$-term. For example, the $t$-term with a linear $d$ in 
${\rm H}_{1}$ would involve quasiparticle pairing between {\it different} bands which is a 
{\it high energy} process now. Subsequently, the pairing between quasiparticles would only
come from the four-fermion attraction in the $J$-term of the Hamiltonian. If we still use the 
similar mean-field ansatz to treat this Hamiltonian, the results would be incorrect because the 
compressibility is always negative. This implies that the solution is thermodynamically unstable. 
Such an instability is actually similar to the spiral instability\cite{ss} in 2D case when one tries to 
generalize the Schwinger-boson mean-field theory to the doped case without considering the phase 
string effect\cite{Weng98}. Therefore, it is necessary to take into account the phase string effect in 
order to acquire a correct mean-field theory on the doped antiferromagnets regardless of 
dimensionality.

M\"{u}ller and Rice have suggested a possible $C2S2$ phase existing between the Nogaoka 
phase with very small vaules of $J/t$ and the $C1S0$ phase with intermediate values of 
$J/t$\cite{Muller}. They provided some numerical evidence to support this conjecture. 
Physically, the appearence of this phase can be understood as the following: When the holons 
move fast, i.e. $t \gg J, J_{\perp}$, the gain in kinetic energy may outweights the cost by 
breaking the rung hole pairs and rung singlets. Thus, the phase coherence between bonding 
and anti-bonding bands is lost and the two-leg ladder is effectively decoupled into two chains 
at low energy. If this picture is correct, then the bond operators (especially the $d$-operator) 
are no longer a good description of the low energy degrees of freedom in this region. On the 
other hand, as we have seen in eq. (\ref{H1e}), there are always attractions between 
quasiparticles on the nearest-neighbor sites arising from breaking the singlet bonds. If they are 
not completely compensated by some repulsive forces at least at the intermediate scale, the 
underlying magnetic structure may still be a gapped spin liquid with a small spin gap and the 
observed $C2S2$ phase perhaps is a finite size effect. But the present mean-field theory, which 
works in the limit $J_{\perp} \gg J,t$, cannot be directly applied to this regime to address those
issues. 

For large doping concentration, the anti-bonding band of electrons is empty and the system falls 
into the $C1S1$ phase. To describe this phase, the bond-operator representation is not 
convenient. We have to go back to eq. (\ref{tJ2}). Nevertheless, the problem that there are 
non-trivial phase factors in the Hamiltonian rears its head again. These phase factors are the 
interactions arising from the phase string effect and entail careful treatment. Otherwise, important 
physics may be lost. The pursuit along this direction is beyond the scope of the present paper.

%%%%%%%%%%%%%%%%%%%%%%%%%%%%%%%%%%%%%%%%%%%%%%%%%
%\begin{center}
% {\Large \bf Acknowledgements}
%\end{center}
\acknowledgements

We are very grateful to T.K. Lee, D.N. Sheng, C.S.Ting, B. Friedman, and D.J. Scalapino for 
stimulating discussions. We also wish to thank the National Center for Theoretical Science of 
National Science Council of R.O.C. for organizing the ''Strongly Correlated Electron Systems'' 
topical program, where this work was initiated. Y.L.L. would like to thank H.L. Lai for 
resolving the problem of programming. The work of Y.L.L. and Y.W.L. is supported by 
National Science Council of ROC under Grant No. NSC88-2811-M-007-0027.  C.-Y.M. is 
supported by National Science Council of ROC under Grant No. NSC88-2112-M-007-009. 
Z.Y.W. acknowledges the support by the Texas ARP program No. 3652707 and the State of 
Texas through the Texas Center for Superconductivity at University of Houston.

%%%%%%%%%%%%%%%%%%%%%%%%%%%%%%%%%%%%%%%%%%%%%%%%
\appendix

\section{The bond-operator representations of bilinear operators in the Hamiltonian}

In this Appendix, we list the bond-operator representations of those bilinear operators 
appearing in the Hamiltonian in the following:
\begin{eqnarray}
h^{+}_{j,1}b_{j,1,\sigma} &=& -\frac{(-\sigma)^{j+1}}{\sqrt{2}}%
a^{+}_{j,-\sigma}(s_{j} +\sigma
t_{j,0})+(-\sigma)^{j+1}a^{+}_{j,\sigma}t_{j,\sigma}+ (-\sigma)^{j}d^{+}_{j}%
\bar{a}_{j,\sigma},  \nonumber \\
h^{+}_{j,2}b_{j,2,\sigma} &=& \frac{(-\sigma)^{j}}{\sqrt{2}}\bar{a}%
^{+}_{j,-\sigma}(s_{j} -\sigma t_{j,0})+(-\sigma)^{j}\bar{a}%
^{+}_{j,\sigma}t_{j,\sigma}+ (-\sigma)^{j+1}d^{+}_{j}a_{j,\sigma},  \nonumber
\\
\sum_{\sigma}b_{j,1,\sigma}b_{j,2,-\sigma} &=& (-1)^{j}\sqrt{2}s_{j}, 
\nonumber \\
b^{+}_{j,1,\sigma}b_{j,1,\sigma} &=& \frac{1}{2}\{s^{+}_{j}s_{j}+\sum_{%
\alpha = \pm 1, 0} t^{+}_{j,\alpha}t_{j,\alpha}+\sigma
\sum_{\sigma^{\prime}=\pm 1}
\sigma^{\prime}t^{+}_{j,\sigma^{\prime}}t_{j,\sigma^{\prime}}+\sigma
(s^{+}_{j}t_{j,0}+t^{+}_{j,0}s_{j})\}  \nonumber \\
& & +\bar{a}^{+}_{j,\sigma}\bar{a}_{j,\sigma},  \nonumber \\
b^{+}_{j,2,\sigma}b_{j,2,\sigma} &=& \frac{1}{2}\{s^{+}_{j}s_{j}+\sum_{%
\alpha = \pm 1, 0} t^{+}_{j,\alpha}t_{j,\alpha}+\sigma
\sum_{\sigma^{\prime}=\pm 1}
\sigma^{\prime}t^{+}_{j,\sigma^{\prime}}t_{j,\sigma^{\prime}}-\sigma
(s^{+}_{j}t_{j,0}+t^{+}_{j,0}s_{j})\}  \nonumber \\
& & +a^{+}_{j,\sigma}a_{j,\sigma},  \nonumber \\
b^{+}_{j,1,\sigma}b_{j,1,-\sigma} &=& \frac{(-1)^{j}}{\sqrt{2}}%
\{t^{+}_{j,\sigma}(s_{j}- \sigma t_{j,0})+(s^{+}_{j}+\sigma
t^{+}_{j,0})t_{j,-\sigma}\}  \nonumber \\
& & +(-1)^{j}\bar{a}^{+}_{j,\sigma}\bar{a}_{j,-\sigma},  \nonumber \\
b^{+}_{j,2,\sigma}b_{j,2,-\sigma} &=& \frac{(-1)^{j}}{\sqrt{2}}%
\{t^{+}_{j,\sigma}(s_{j}+ \sigma t_{j,0})+(s^{+}_{j}-\sigma
t^{+}_{j,0})t_{j,-\sigma}\}  \nonumber \\
& & +(-1)^{j+1}a^{+}_{j,\sigma}a_{j,-\sigma}.  \label{hbbb}
\end{eqnarray}
The above equations should be considered only in the physical Hilbert space.

%%%%%%%%%%%%%%%%%%%%%%%%%%%%%%%%%%%%%%%%%%
\section{The form of ${\rm H}_{2}$}

We list ${\rm H}_{2}$ in the following: 
\begin{eqnarray}
{\rm H}_{2} &=& \frac{t}{2}\sum_{j}\sigma
(s_{j+1}t^{+}_{j,0}+t_{j+1,0}s^{+}_{j}) (a^{+}_{j+1,\sigma}a_{j,\sigma}-\bar{%
a}^{+}_{j+1,\sigma} \bar{a}_{j,\sigma})  \nonumber \\
& & +i\frac{t}{\sqrt{2}}\sum_{j}\sigma (s_{j+1}t^{+}_{j,-\sigma}
-t_{j+1,\sigma}s^{+}_{j})(a^{+}_{j+1,\sigma}a_{j,-\sigma}+ \bar{a}%
^{+}_{j+1,\sigma}\bar{a}_{j,-\sigma})  \nonumber \\
& & -i\frac{t}{\sqrt{2}}\sum_{j}(t_{j+1,0}t^{+}_{j,-\sigma}+
t_{j+1,\sigma}t^{+}_{j,0})(a^{+}_{j+1,\sigma}a_{j,-\sigma}- \bar{a}%
^{+}_{j+1,\sigma}\bar{a}_{j,-\sigma})  \nonumber \\
& & -\frac{t}{\sqrt{2}}\sum_{j}(t_{j+1,0}d_{j}+d_{j+1}t_{j,0})
(a^{+}_{j+1,\sigma}\bar{a}^{+}_{j,-\sigma} +\bar{a}^{+}_{j+1,%
\sigma}a^{+}_{j,-\sigma})  \nonumber \\
& & -it \sum_{j}(t_{j+1,\sigma}d_{j}-d_{j+1}t_{j,\sigma}) (a^{+}_{j+1,\sigma}%
\bar{a}^{+}_{j,\sigma} -\bar{a}^{+}_{j+1,\sigma}a^{+}_{j,\sigma})  \nonumber
\\
& & -\frac{J}{8}\sum_{j}\sigma \{(s_{j}t^{+}_{j,0}+t_{j,0}s^{+}_{j})
(a^{+}_{j+1,\sigma}a_{j+1,\sigma}-\bar{a}^{+}_{j+1,\sigma} \bar{a}%
_{j+1,\sigma})+(j \leftrightarrow j+1)\}  \nonumber \\
& & +\frac{J}{8}\sum_{j}\sigma \{(\sum_{\sigma^{\prime}=\pm 1}
\sigma^{\prime}t^{+}_{j,\sigma^{\prime}}t_{j,\sigma^{\prime}})
(a^{+}_{j+1,\sigma}a_{j+1,\sigma}+\bar{a}^{+}_{j+1,\sigma} \bar{a}%
_{j+1,\sigma})+(j \leftrightarrow j+1)\}  \nonumber \\
& & -\frac{J}{2}\sum_{j,\alpha=\pm 1,0}(t_{j,0}t_{j+1,0}t^{+}_{j,\alpha}
t^{+}_{j+1,-\alpha}-t_{j,0}t^{+}_{j+1,0}t^{+}_{j,\alpha}t_{j+1,\alpha}) 
\nonumber \\
& & -i\frac{J}{2\sqrt{2}}\sum_{j}\sigma \{a^{+}_{j+1,\sigma}
a_{j+1,-\sigma}t^{+}_{j,-\sigma}(s_{j}-\sigma t_{j,0})+ \bar{a}%
^{+}_{j+1,\sigma}\bar{a}_{j+1,-\sigma}t^{+}_{j,-\sigma}(s_{j}+ \sigma
t_{j,0})  \nonumber \\
& & -(j \leftrightarrow j+1)\}  \nonumber \\
& & +\frac{J}{4}\sum_{j}(\sum_{\sigma =\pm 1}t^{+}_{j,\sigma}t_{j,\sigma})
(\sum_{\sigma^{\prime}=\pm 1}t^{+}_{j+1,\sigma^{\prime}}
t_{j+1,\sigma^{\prime}})+H.c. .
\end{eqnarray}

%%%%%%%%%%%%%%%%%%%%%%%%%%%%%%%%%%%%%%%%%%%%%
\section{$\Delta$ and $\bar{\Delta}$ are spin singlets}

We would like to discuss the transformation properties of $\Delta$ and $\bar{\Delta}$ under 
the spin rotation and verify that they are spin singlets. Due to the same reason as that for $t$ 
operators, $a_{\sigma}$ and $\bar{a}_{\sigma}$ defined in (\ref{bond}) are not spinors. 
Therefore, we are unable to directly conclude that $\Delta$ and $\bar{\Delta}$ are not tensors 
under the spin rotation. To make the rotational symmetry manifest, we perform the following 
unitary transformation: 
\begin{eqnarray}
a^{+}_{j, \sigma} &\rightarrow& a^{+}_{j, \sigma}\exp{\{-i\sigma \frac{\pi}{4%
}-i\sigma \frac{\pi}{2} \sum_{l < j}(n^{h}_{l,1}+n^{h}_{l,2})\}},  \nonumber
\\
\bar{a}^{+}_{j, \sigma} &\rightarrow& \bar{a}^{+}_{j, \sigma}\exp{\{i\sigma 
\frac{\pi}{4} -i\sigma \frac{\pi}{2}\sum_{l < j}(n^{h}_{l,1}+n^{h}_{l,2})\}}.
\label{atran}
\end{eqnarray}
Under the above transformation (\ref{atran}), $\Delta+\bar{\Delta}$ and 
$\Delta-\bar{\Delta}$ become 
\begin{eqnarray}
\Delta+\bar{\Delta} &\rightarrow& i\sum_{\sigma}\sigma \langle
a^{+}_{j+1,\sigma} a^{+}_{j,-\sigma}-\bar{a}^{+}_{j+1,\sigma}\bar{a}%
^{+}_{j,-\sigma}\rangle ,  \nonumber \\
\Delta-\bar{\Delta} &\rightarrow& \sum_{\sigma}\sigma \langle
a^{+}_{j+1,\sigma} \bar{a}^{+}_{j,-\sigma}+\bar{a}^{+}_{j+1,%
\sigma}a^{+}_{j,-\sigma}\rangle . 
\end{eqnarray}
Now it is clear that $\Delta$ and $\bar{\Delta}$ are indeed spin singlets. We also examine 
other terms in ${\rm H}$ with eq. (\ref{atran}) and confirm that our mean-field ansatz 
respects the rotational symmetry. 

%%%%%%%%%%%%%%%%%%%%%%%%%%%%%%%%%%%%%%%%%

\section{Diagonalization of eq.(\ref{HMF})}

Here we present the procedure to diagonalize eq.(\ref{HMF}). After performing 
Fourier transformations on all operators by 
$\hat{O}_{j}=\frac{1}{\sqrt{N}}\sum_{k} \hat{O}_{k}e^{ikx_{j}}$, the 
mean-field Hamiltonian in eq.(\ref{HMF}) becomes 
\begin{eqnarray}
{\rm H}_{MF} &=& N\{\lambda -\frac{J_{\perp}}{4}-\frac{J}{2}-(\frac{3}{4}%
J_{\perp}+ \lambda)\bar{s}^{2}+(\chi+\bar{\chi}) Qt+ \frac{J}{8}(\Delta+\bar{%
\Delta})^{2}  \nonumber \\
& & -\frac{J}{16}(\chi -\bar{\chi})^{2}-(\frac{J}{2}\bar{d}^{2}+2\mu
+\lambda -\frac{J_{\perp}}{4}-J)\bar{d}^{2}\}  \nonumber \\
& & +\sum_{k}(\epsilon_{k}e^{+}_{k\sigma}e_{k\sigma}+\bar{\epsilon}_{k} \bar{%
e}^{+}_{k\sigma}\bar{e}_{k\sigma})  \nonumber \\
& & +\sum_{k}(i\Gamma_{k}e^{+}_{k\uparrow}e^{+}_{-k\downarrow}+ i\bar{\Gamma}%
_{k}\bar{e}^{+}_{k\uparrow}\bar{e}^{+}_{-k\downarrow}+ H.c.)  \nonumber \\
& & +\sum_{k}\{\Lambda_{k}t^{+}_{k\alpha}t_{k\alpha}+\Pi_{k}
(t^{+}_{k\alpha}t^{+}_{-k-\alpha}+t_{k\alpha}t_{-k-\alpha})\}
\label{MMF}  
\end{eqnarray}
where 
\begin{eqnarray}
\epsilon_{k} &=& -(t(2\bar{d}^{2}+\bar{s}^{2}+Q)-\frac{J}{8}(\chi -\bar{\chi}%
))\cos{k}  \nonumber \\
& & -t-\mu -\lambda +\frac{J_{\perp}}{4}+\frac{J}{2}(1-\bar{d}^{2}), 
\nonumber \\
\bar{\epsilon}_{k} &=& -(t(2\bar{d}^{2}+\bar{s}^{2}+Q)+\frac{J}{8}(\chi -%
\bar{\chi}))\cos{k}  \nonumber \\
& & +t-\mu -\lambda +\frac{J_{\perp}}{4}+\frac{J}{2}(1-\bar{d}^{2}), 
\nonumber \\
\Gamma_{k} &=& (\frac{J}{4}(\Delta+\bar{\Delta})+2\sqrt{2}t\bar{s}\bar{d}%
)\sin{k},   \nonumber \\
\bar{\Gamma}_{k} &=& (\frac{J}{4}(\Delta+\bar{\Delta})-2\sqrt{2}t\bar{s}\bar{%
d})\sin{k},  \nonumber \\
\Lambda_{k} &=& (J\bar{s}^{2}-t(\chi+\bar{\chi}))\cos{k}+\frac{J_{\perp}}{4}%
-\lambda ,  \nonumber \\
\Pi_{k} &=& \frac{J}{2}\bar{s}^{2}\cos{k}.  \nonumber
\end{eqnarray}
Here the lattice spacing has been taken to be unity.

Eq. (\ref{MMF}) can be diagonalized by the following Bogolioubov transformations: 
\begin{eqnarray}
\alpha_{k} &=& u_{k}e_{k\uparrow}-v_{k}e^{+}_{-k\downarrow}, \ \beta_{k} =
u_{k}e_{-k\downarrow}+v_{k}e^{+}_{k\uparrow},  \label{be1} \\
\bar{\alpha}_{k} &=& \bar{u}_{k}\bar{e}_{k\uparrow}-\bar{v}_{k}\bar{e}%
^{+}_{-k\downarrow}, \ \bar{\beta}_{k} = \bar{u}_{k}\bar{e}_{-k\downarrow}+%
\bar{v}_{k}\bar{e}^{+}_{k\uparrow},  \label{be2} \\
\gamma_{k\alpha} &=& \cosh{\theta_{k}}t_{k\alpha}+\sinh{\theta_{k}}%
t^{+}_{-k-\alpha}.  \label{bt}
\end{eqnarray}
The coefficients $u_{k}$, $v_{k}$, $\bar{u}_{k}$, $\bar{v}_{k}$, $\cosh{%
\theta_{k}}$, and $\sinh{\theta_{k}}$ are given by 
\begin{eqnarray}
u_{k} &=& \cos{\phi_{k}}e^{i\frac{\pi}{4}}, \ v_{k} = \sin{\phi_{k}}e^{-i%
\frac{\pi}{4}},  \nonumber \\
\bar{u}_{k} &=& \cos{\bar{\phi}_{k}}e^{i\frac{\pi}{4}}, \ \bar{v}_{k} = \sin{%
\bar{\phi}_{k}} e^{-i\frac{\pi}{4}},  \nonumber \\
\cos^{2}{\phi_{k}} &=& \frac{1}{2}(1+\frac{\epsilon_{k}}{E_{k}}) , \ \sin^{2}%
{\phi_{k}} = \frac{1}{2}(1-\frac{\epsilon_{k}}{E_{k}}) ,  \nonumber \\
\cos^{2}{\bar{\phi}_{k}} &=& \frac{1}{2}(1+\frac{\bar{\epsilon}_{k}}{\bar{E}%
_{k}}) , \ \sin^{2}{\bar{\phi}_{k}} = \frac{1}{2}(1-\frac{\bar{\epsilon}_{k}%
}{\bar{E}_{k}}) ,  \nonumber \\
E_{k} &=& \sqrt{\epsilon^{2}_{k}+\Gamma^{2}_{k}}, \ \bar{E}_{k} = \sqrt{\bar{%
\epsilon}^{2}_{k}+\bar{\Gamma}^{2}_{k}},  \label{uuvv} \\
\cosh^{2}{\theta_{k}} &=& \frac{1}{2}(\frac{\Lambda_{k}}{\omega_{k}}+1), \
\sinh^{2}{\theta_{k}} = \frac{1}{2}(\frac{\Lambda_{k}}{\omega_{k}}-1), 
\nonumber \\
\omega_{k} &=& \sqrt{\Lambda^{2}_{k}-4\Pi^{2}_{k}}.  \label{gam}
\end{eqnarray}
In eq. (\ref{gam}), both $\Lambda_{k}$ and $\Lambda^{2}_{k}-4\Pi^{2}_{k}$ 
have to be positive. This constraint will be enforced in our numerical analysis. If 
we define $\bar{\nu}=(\chi+\bar{\chi}) t/(\frac{J_{\perp}}{4}-\lambda)$
and $\nu = 2J\bar{s}^{2}/(\frac{J_{\perp}}{4}-\lambda)$, then the band 
minimum of $t$ particles occurs at $k=\pi$ when $\bar{\nu}\leq \nu /2$ and at 
$k=0$ when $\bar{\nu}>\nu /2$. With the help of eqs. (\ref{be1}), (\ref{be2}), 
and (\ref{bt}), we obtain eq. (\ref{HMF3}).  

%%%%%%%%%%%%%%%%%%%%%%%%%%%%%%%%%%%%%%%%%%%%

\section{The bond-operator representations of pairing fields}

Here we give the bond-operator representations of pairing fields in the phase
string formulae. First, we need the corresponding representation of electron
operators. They are as the following: 
\begin{eqnarray}
c_{j,1,\sigma} &=& \exp{(-i\frac{\pi}{4}(1+\sigma))}\exp{\{i\frac{\pi}{2}%
\sum_{l>j,m} (2S^{z}_{l,m}-1-\sigma n^{h}_{l,m})\}}  \nonumber \\
& & \{\frac{\sigma}{\sqrt{2}}a^{+}_{j,-\sigma}(s_{j}+\sigma t_{j,0})
-ia^{+}_{j,\sigma}t_{j,\sigma}+d^{+}_{j}\bar{a}_{j,\sigma}\} \sigma^{N_{h}},
\nonumber \\
c_{j,2,\sigma} &=& \exp{(i\frac{\pi}{4}(1+\sigma))}\exp{\{i\frac{\pi}{2}%
\sum_{l>j,m} (2S^{z}_{l,m}-1-\sigma n^{h}_{l,m})\}}  \nonumber \\
& & \{-\frac{\sigma}{\sqrt{2}}\bar{a}^{+}_{j,-\sigma}(s_{j}-\sigma t_{j,0})
+i\bar{a}^{+}_{j,\sigma}t_{j,\sigma}+d^{+}_{j}a_{j,\sigma}\} \sigma^{N_{h}}.
\label{cb}
\end{eqnarray}
By pluging the above formulae into eq. (\ref{pf}), we obtain 
\begin{eqnarray}
\Delta_{y}(j) &=& -\frac{i}{2}d^{+}_{j}s_{j}\exp{(2\pi
i\sum_{l>j,m}S^{z}_{l,m})},  \nonumber \\
\Delta_{x}(j) &=& \exp{(2\pi i\sum_{l>j+1,m}S^{z}_{l,m})}  \nonumber \\
& & \{\frac{i}{\sqrt{2}}(d^{+}_{j+1}d^{+}_{j}a_{j+1,\sigma}a_{j,-\sigma} -%
\bar{a}^{+}_{j+1,\sigma}\bar{a}^{+}_{j,-\sigma}t_{j+1,\sigma}t_{j,-\sigma}) 
\nonumber \\
& & -\frac{1}{\sqrt{2}}(d^{+}_{j}t_{j+1,\sigma}\bar{a}^{+}_{j+1,%
\sigma}a_{j,-\sigma} +d^{+}_{j+1}t_{j,-\sigma}a_{j+1,\sigma}\bar{a}%
^{+}_{j,-\sigma})  \nonumber \\
& & +\frac{i}{2}\sigma [d^{+}_{j+1}a_{j+1,\sigma}\bar{a}^{+}_{j,\sigma}
(s_{j}+\sigma t_{j,0})-d^{+}_{j}\bar{a}^{+}_{j+1,\sigma}a_{j,\sigma}
(s_{j+1}+\sigma t_{j+1,0})]  \nonumber \\
& & +\frac{\sigma}{2}\bar{a}^{+}_{j+1,\sigma}\bar{a}^{+}_{j,\sigma}[t_{j,%
\sigma} (s_{j+1}+\sigma t_{j+1,0})-t_{j+1,\sigma}(s_{j}+\sigma t_{j,0})] 
\nonumber \\
& & +\frac{i}{2\sqrt{2}}\bar{a}^{+}_{j+1,\sigma}\bar{a}^{+}_{j,-\sigma}
(s_{j+1}+\sigma t_{j+1,0})(s_{j}-\sigma t_{j,0})\}.
\end{eqnarray}
We have to emphasize that it is necessary to take into account $\sigma^{N_{h}}$ in 
eq. (\ref{cb}). Otherwise, the representations of pairing fields would be incorrect.

%%%%%%%%%%%%%%%%%%%%%%%%%%%%%%%%%%%%%%%%%%%%%

\newpage

\begin{center}
{\large {\bf Figure Captions}}:
\end{center}

\begin{description}
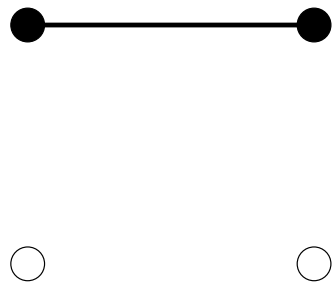
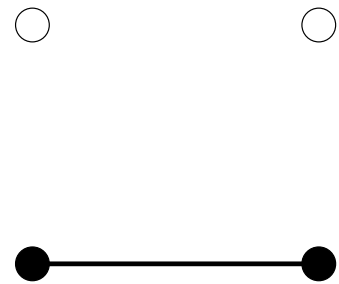
\item[Fig. 1]  The local structure of hole pairs: (a) diagonal site. (b)
chain direction. The solid line and open circles represent the spin singlet
bonds and holes, respectively.

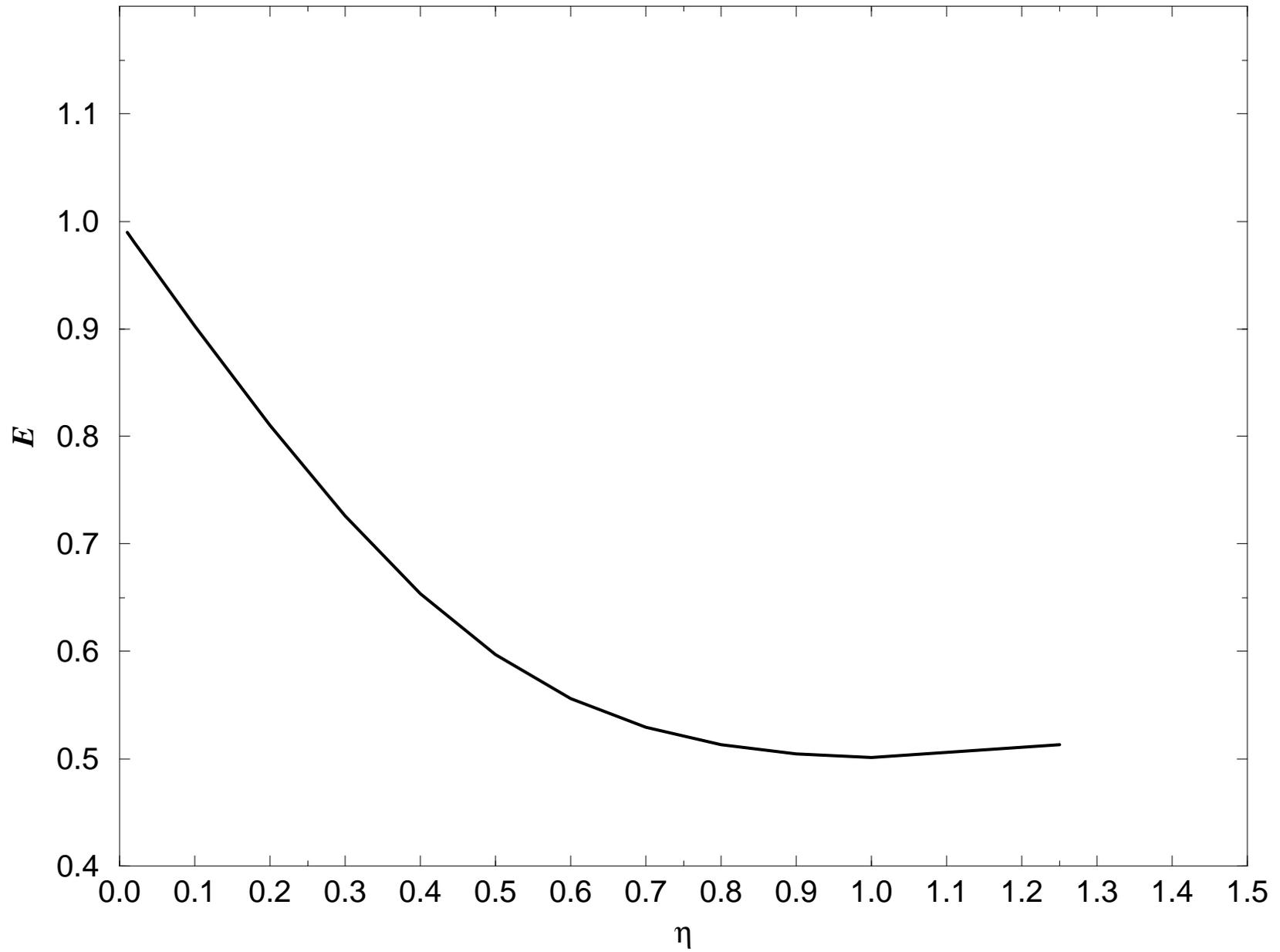
\item[Fig. 2]  The spin gap, $\Delta _t$, as a function of $\eta =J/J_{\perp }$ 
where $E=\Delta _t/J_{\perp }$.

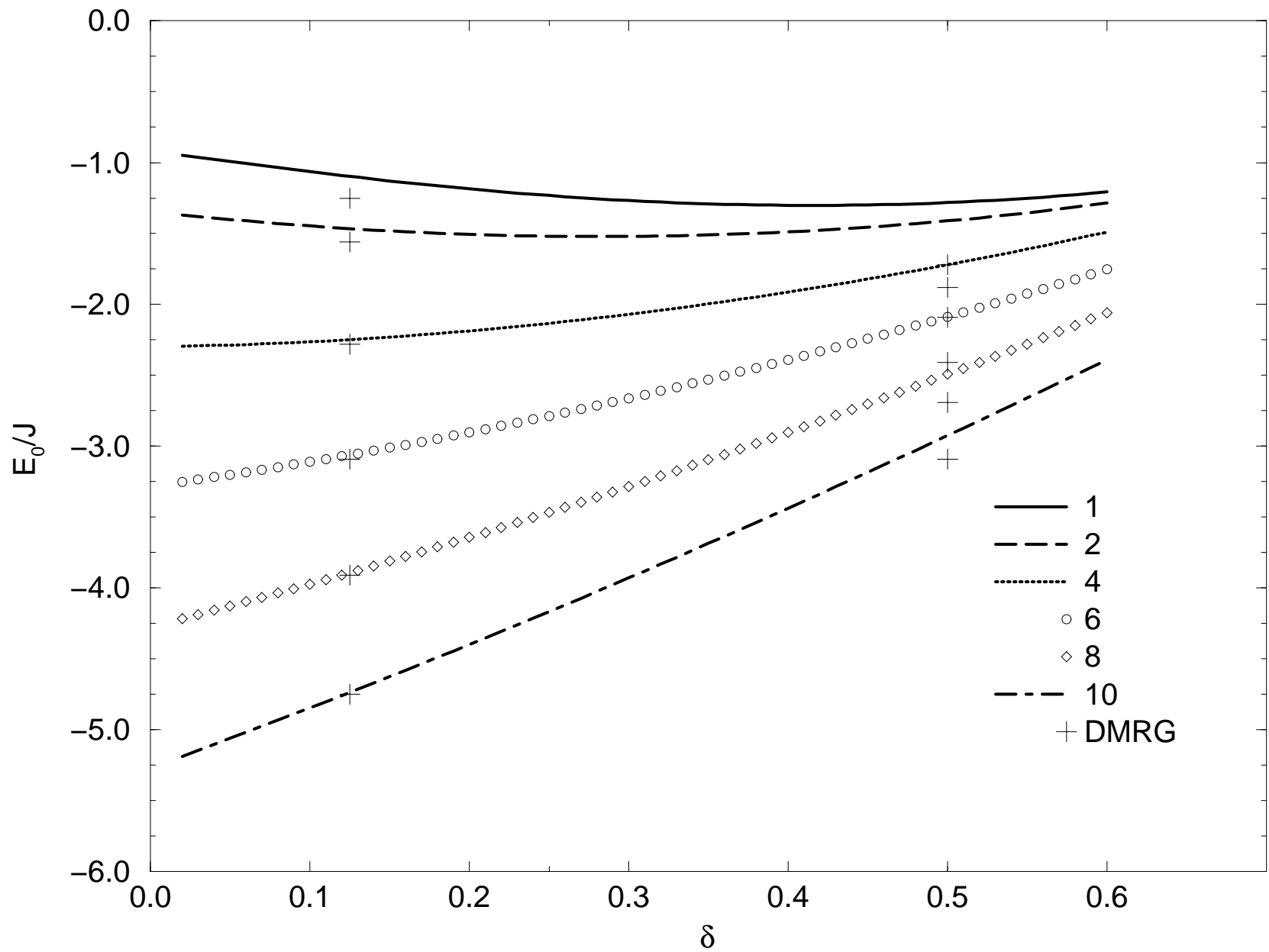
\item[Fig. 3] The ground state energy per site with $J/t=0.5$. Different 
curves correspond to $J_{\perp}/J=1, 2, 4, 6, 8$, and $10$. The data denoted 
by the cross are obtained with DMRG and taken from Ref.\cite{Sierra97}.

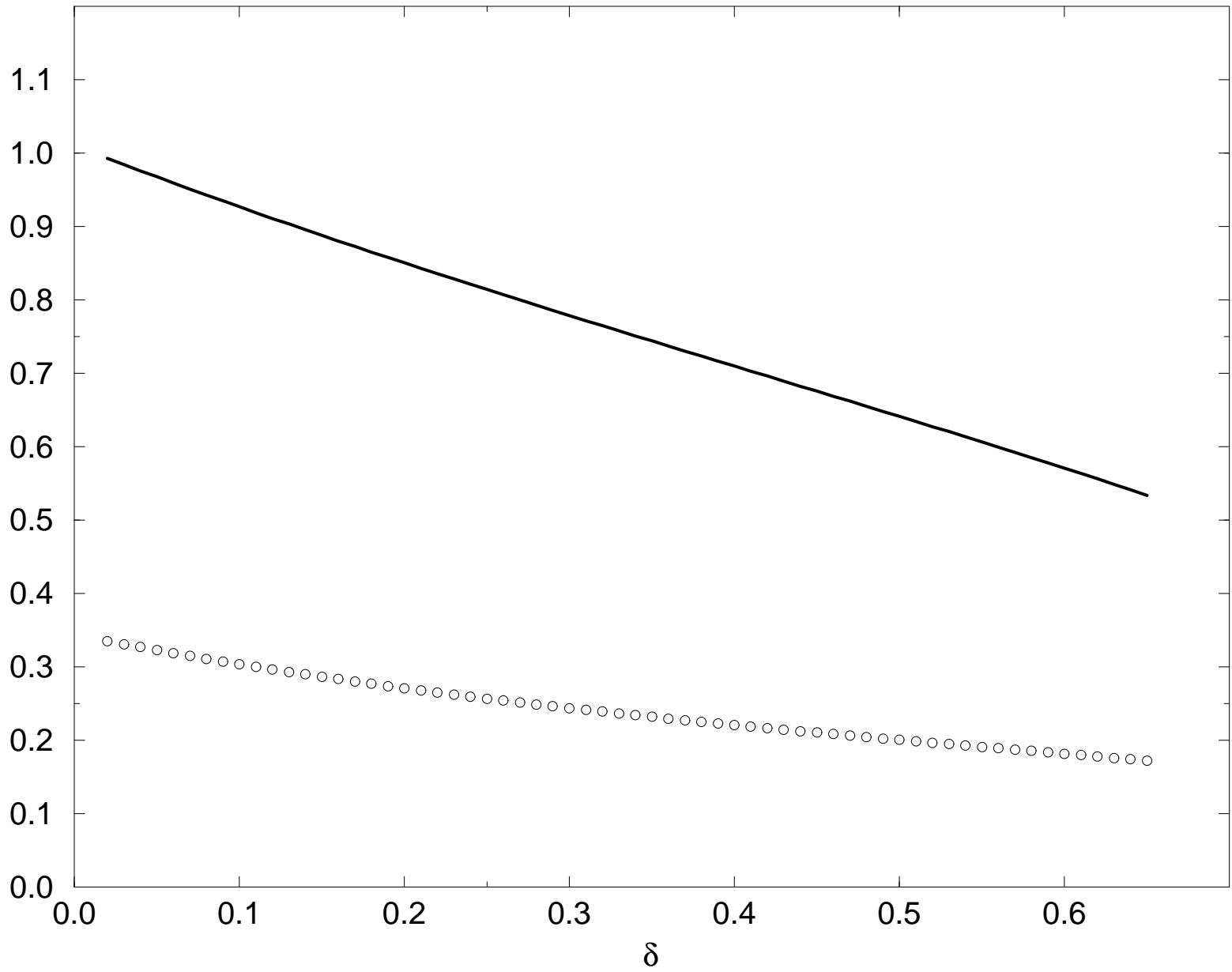
\item[Fig. 4]  The weight of different types of hole pairs with $J/t=0.5$ and 
$J_{\perp }/J=10$. The solid line and open circles represent the weights of 
hole pairs on diagonal sites and chain direction, respectively. 

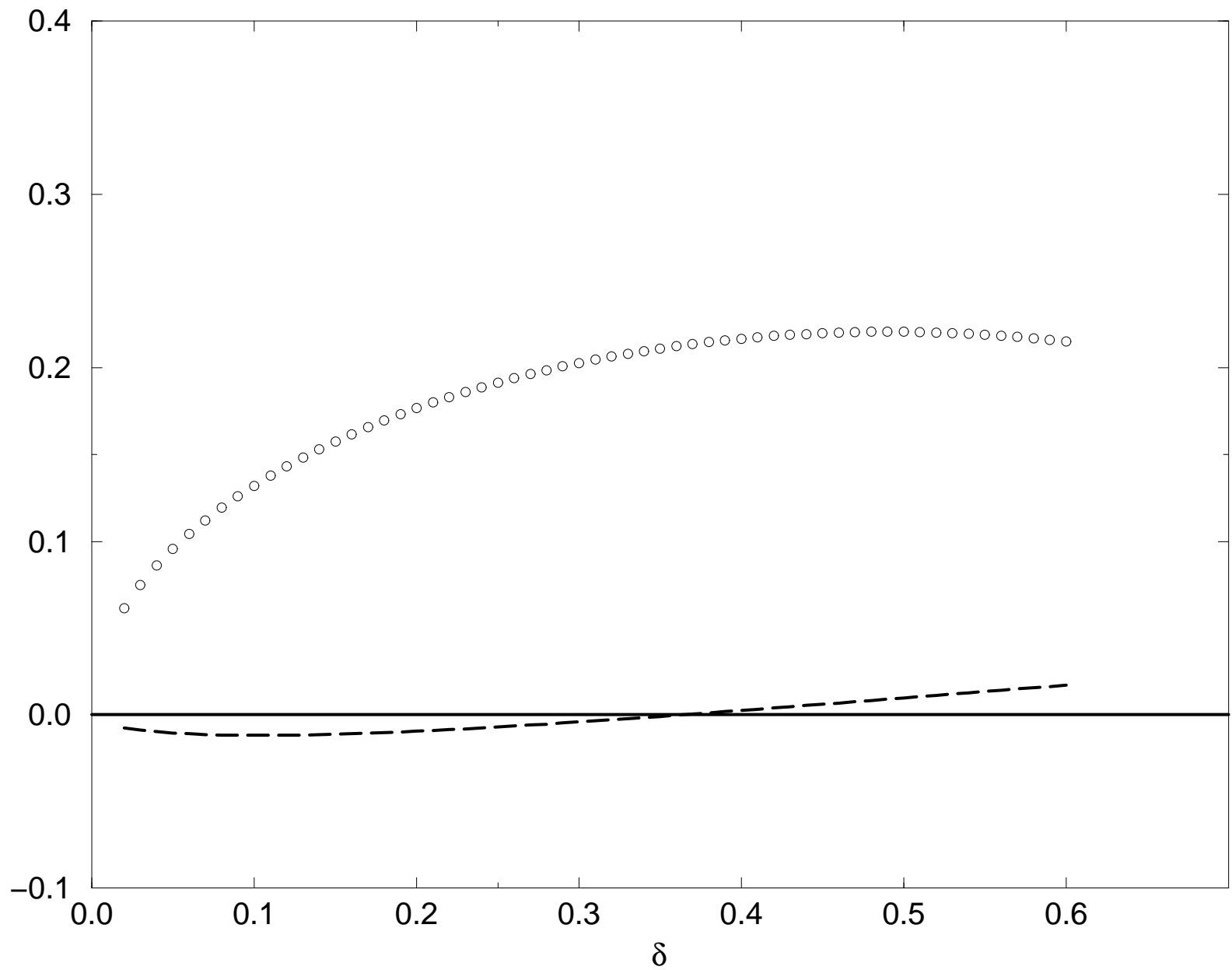
\item[Fig. 5]  The vacuum expectation values of pairing fields as functions of 
the doping concentration at $J/t=0.5$ and $J_{\perp }/J=10$. The dashed line 
and open circles represent $\Delta _x$ and $\Delta _y$, respectively. 

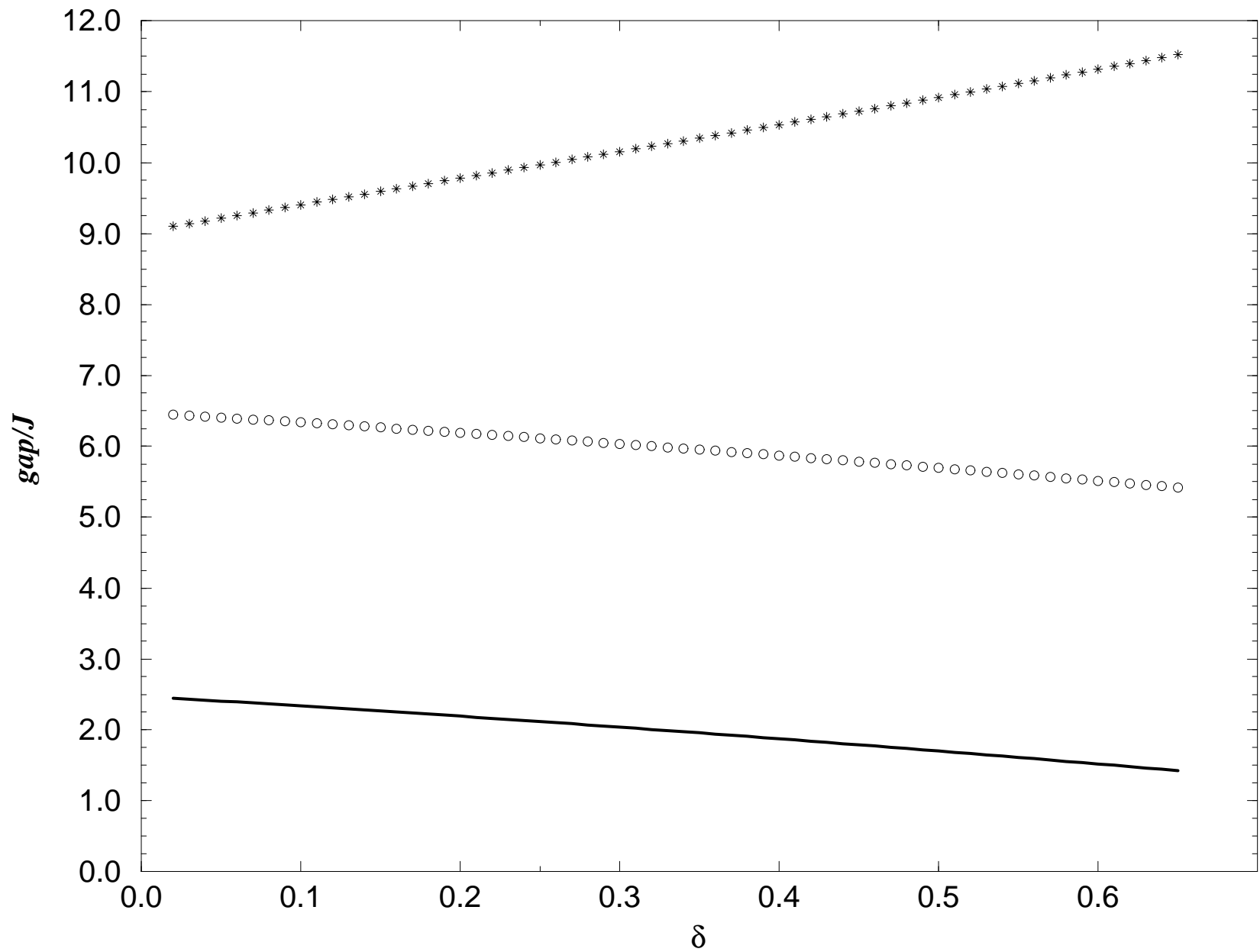
\item[Fig. 6]  The gaps of spin excitations at $J/t=0.5$ and $J_{\perp }/J=10$. 
The solid line, open circles, and stars correspond to quasiparticles in bonding 
band, anti-bonding band, and magnons, respectively. 

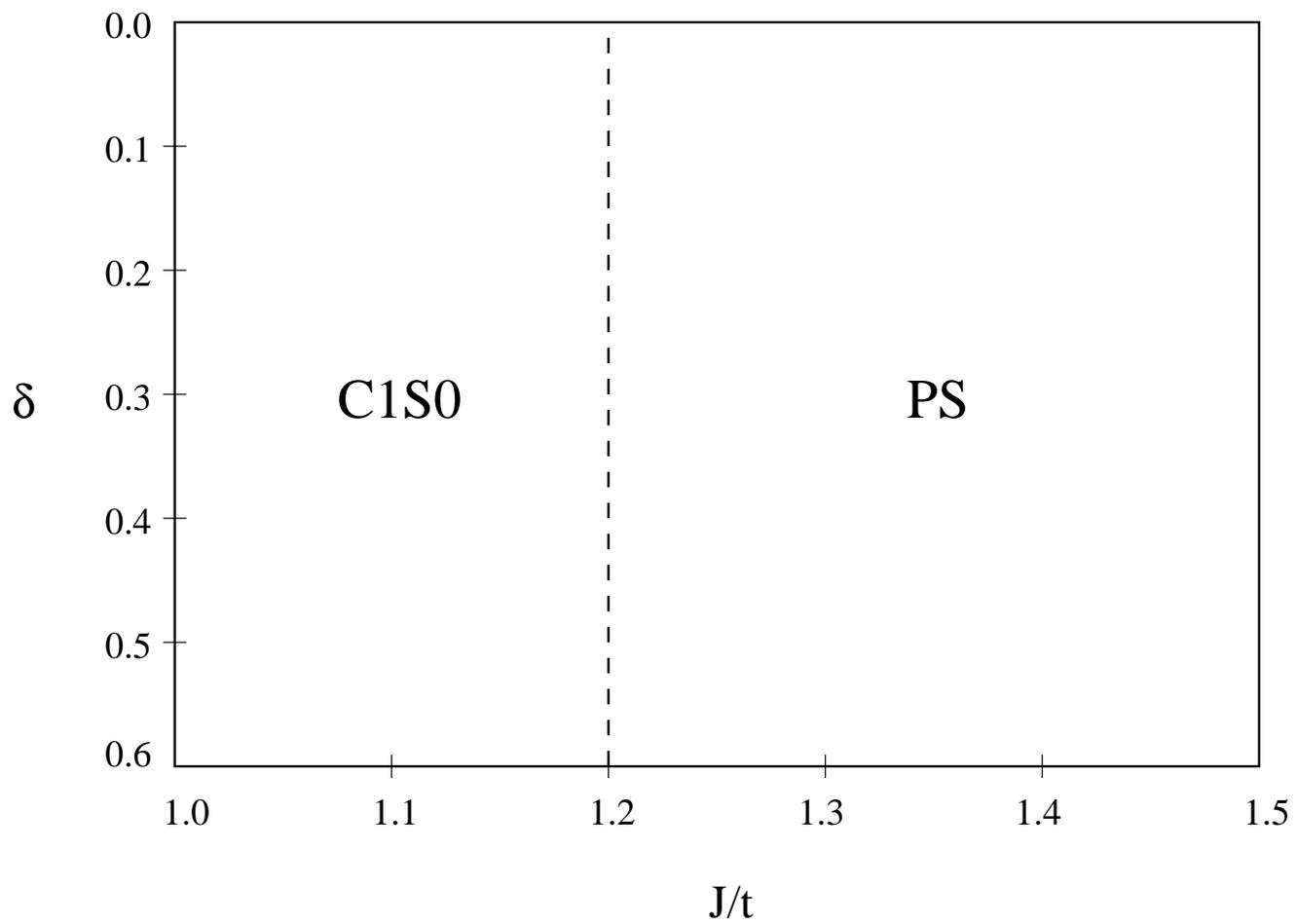
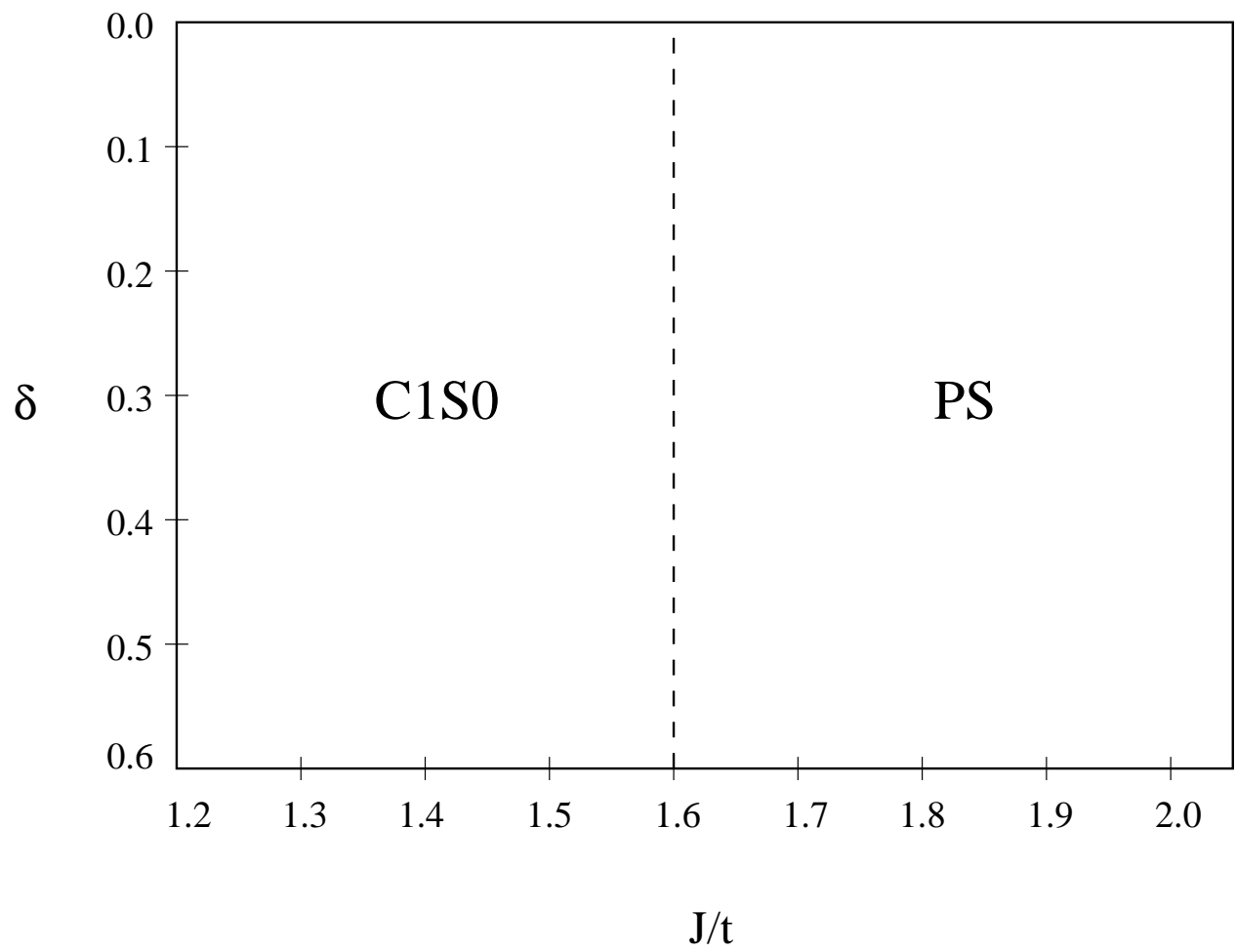
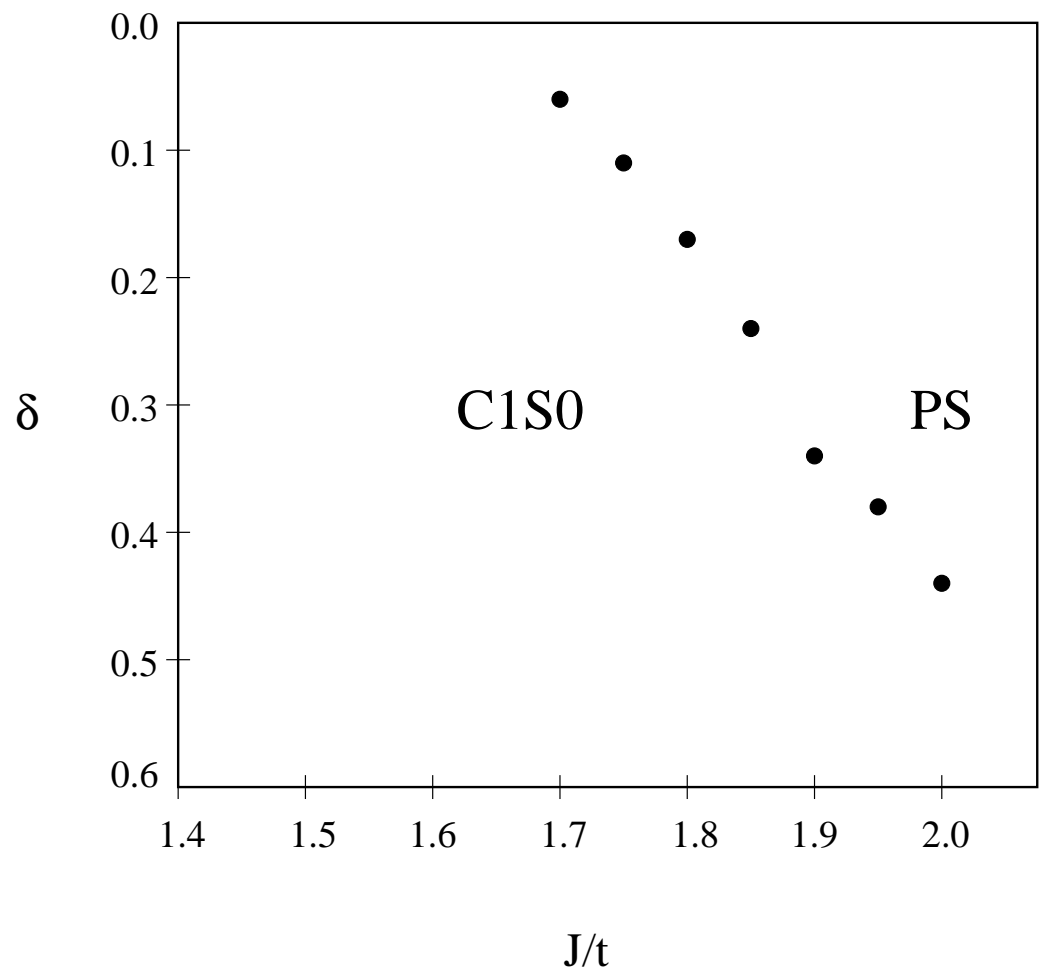
\item[Fig. 7] The boundary between the $C1S0$ phase and the phase separation 
region with (a) $J_{\perp}/J=10$, (b) $J_{\perp}/J=5$, and (c) $J_{\perp}/J=2$.

\end{description}


\begin{references}
\bibitem{Dago96}  For a review, see E. Dagotto and T.M. Rice, Science {\bf %
271}, 618 (1996).

\bibitem{Dagotto92}  E. Dagotto, J. Riera, and D. Scalapino, Phys. Rev. {\bf %
B 45}, 5744 (1992); T.M. Rice, S. Gopalan, and M. Sigrist, Europhys. Lett. 
{\bf 23}, 445 (1993).

\bibitem{Barnes}  T. Barnes, E. Dagotto, J. Riera and E.S. Swanson, Phys.
Rev. {\bf B 47} , 3196 (1993).

\bibitem{Reig}  M. Reigrotzki, H. Tsunetsugu and T.M. Rice, J. Phys:
Condens. Matter {\bf 6}, 9235 (1994).

\bibitem{Gopalan}  S. Gopalan, T. M. Rice, and M. Sigrist, Phys. Rev. {\bf B
49}, 8901 (1994).

\bibitem{Noack}  S.R. White, R.M. Noack, and D.J. Scalapino, 
Phys. Rev. Lett {\bf 73}, 886 (1994).

\bibitem{Sigrist}  M. Sigrist, T.M. Rice, and F.C. Zhang, Phys. Rev. {\bf B
49}, 12058 (1994).

\bibitem{Poil}  D. Poilblanc, D.J. Scalapino, and W. Hanke, Phys. Rev. {\bf %
B 52}, 6796 (1995).

\bibitem{Troyer}  M. Troyer, H. Tsunetsugu, and T.M. Rice, Phys. Rev. {\bf B
53}, 251 (1995); C.A. Hayward and D.Poilblanc, {\it ibid.}, {\bf 53}, 11721
(1996).

\bibitem{White}  S.R. White and D.J. Scalapino, Phys. Rev. {\bf B 55}, 6504
(1997).

\bibitem{Sierra97}  G. Sierra, M.A. Mart\'{i}n-Delgado, J. Dukelsky, S.R.
White, and D.J. Scalapino, Phys. Rev. {\bf B 57}, 11666 (1998).

\bibitem{Muller}  T. F. A. M\"{u}ller and T. M. Rice, Phys. Rev. {\bf B 58},
3425 (1998).

\bibitem{Ammon}  B. Ammon, M. Troyer, T.M. Rice, and N. Shibata, 
cond-mat/9812144.

\bibitem{Azuma}  M. Azuma, Z. Hiroi, M. Takano, K. Ishida, and Y. Kitaoka,
Phys. Rev. Lett. {\bf 73}, 3463 (1994).

\bibitem{Ueha}  M. Uehara, T. Nagata, J. Akimitsu, H. Takahashi, N.
M\^{o}ri, and K. Kinoshita, J. Phys. Soc. Jpn. {\bf 65}, 2764 (1996); T.
Osafune, N. Motoyama, H. Eisaki, and S. Uchida, Phys. Rev. Lett. {\bf 78},
1980 (1997); N. Motoyama, T. Osafune, T. Kakeshita, H. Eisaki, S. Uchida,
Phys. Rev. {\bf B 55}, R3386 (1997).

\bibitem{Maya}  H. Mayaffre, P. Auban-Senzier, M. Nardone, D. J\'{e}rome, D.
Poilblanc, C. Bourbonnais, U. Ammerahl, G. Dhalenne, and A. Revcolevschi,
Science {\bf 279}, 345 (1998).

\bibitem{Imai}  T. Imai, K.R. Thurber, K.M. Shen, A.W. Hunt, and F.C. Chou,
Phys. Rev. Lett. {\bf 81}, 220 (1998); S. Katano, T. Nagata, J. Akimitsu, M. Nishi, 
and K. Kakurai, {\it ibid.}, {\bf 82}, 636 (1999).

\bibitem{Sachdev}  S. Sachdev and R.N. Bhatt, Phys. Rev. {\bf B 41}, 9323
(1990).

\bibitem{Marsh}  W. Marshall, Proc. R. Soc. London Ser. {\bf A 232}, 48
(1955).
                                          
\bibitem{Weng}  Z. Y. Weng, D. N. Sheng, Y.-C. Chen, and C. S. Ting, Phys.
Rev. {\bf B 55}, 3892 (1997).

\bibitem{Weng98}  Z. Y. Weng, D. N. Sheng, and C. S. Ting, Phys. Rev. Lett. 
{\bf 80}, 5401 (1998); Phys. Rev. {\bf B 59}, April 1 (1999).

\bibitem{ss} B.I. Shraiman and E.D. Siggia, Phys. Rev. Lett. {\bf 62}, 
1564 (1989).  

\bibitem{Fisher} L. Balents and M.P.A. Fisher, Phys. Rev. {\bf B 53}, 12133 (1996).

\bibitem{foot4} In the case with periodic boundary conditions or the ring, we 
can form closed paths by winding around it. However, these closed paths only 
give constant phases for fixed hole concentration. They do not affect the 
dynamics of holons and spinons.
  
\bibitem{foot1}  We will not perform the unitary transformation on ${\rm H}_2$. 
For those terms with single $t_{j,\sigma }$ in ${\rm H}_2$, this transformation 
will introduce a nonlocal phase factor, which only depends on the hole density. 
However, these terms are still high energy processes and will not change the 
qualitative behaviors obtained from ${\rm H}_0+{\rm H}_1$.

\bibitem{foot2}  If we do not consider the phase string and apply the bond-operator 
representation to the $t-J$ model directly, then $d$ would be parity odd. As a 
consequence, exchange symmetry forbids the condensation of $d$ particles. However, 
$d$ can condense after taking into account the phase string. This will result in very 
different physics.

\bibitem{Haldane}  See, for example, F.D.M. Haldane, J. Phys. {\bf C 14},
2585 (1981).

\bibitem{Integral}  I.S. Gradshteyn and I.M. Ryzhik, {\it Table of
Integrals, Series, and Products}, 5th edition, (Academic Press, 1994).

\bibitem{foot3}  Of course, there can be gapless excitations at some points
of momentum space if we tune those parameters delicately and this does not
need a vanishing $\Gamma _k$. However, we do not find such a situation.
\end{references}
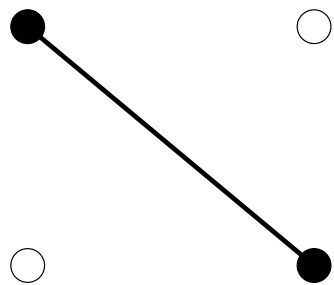
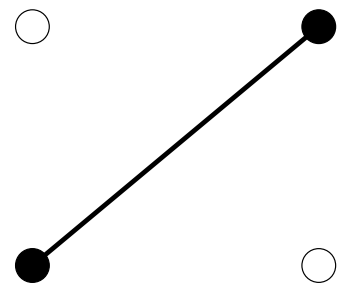
\end{document}